\documentclass[a4paper,12pt]{article}  
\pdfoutput=1
\usepackage{amsmath,amsfonts,amssymb,epsfig,comment,xspace,listings,xcolor,cite}
\usepackage{graphicx,caption,subcaption}
\usepackage[T1]{fontenc}
\numberwithin{equation}{section}
\usepackage{slashed}

\usepackage{youngtab}
\usepackage{listings}
\usepackage{xcolor}
\usepackage{booktabs}

\usepackage{tabularx, multirow}
\usepackage[colorlinks,citecolor=blue,urlcolor=blue,linkcolor=blue]{hyperref}

\makeatletter

\def\pd[#1]{\frac{\partial}{\partial #1}}
\def\pdd[#1,#2]{\frac{\partial #1}{\partial #2}}

\def\beq{\begin{equation}}
\def\eeq{\end{equation}}

\def\twomat[#1,#2][#3,#4]{\left( \begin{array}{cc} #1 & #2 \\ #3 & #4 \end{array} \right)}
\def\twoa[#1,#2][#3,#4]{\left( \begin{array}{cc} #1 & #2 \\ #3 & #4 \end{array} \right)}

\def\thv[#1,#2,#3]{\left( \begin{array}{c} #1 \\ #2 \\ #3 \end{array} \right)}
\def\twv[#1,#2]{\left( \begin{array}{c} #1 \\ #2 \end{array} \right)}

\def\SARAH{{\sc SARAH}\xspace}

\def\MadGraph{{\sc MG5\_aMC}\xspace}
\def\madgraph{\MadGraph}
\def\HackAnalysis{{\sc HackAnalysis}\xspace}
\def\hackanalysis{\HackAnalysis}
\def\hepmc{{\tt hepmc}\xspace}
\def\hepmcv{{\tt hepmc\_v2}\xspace}
\def\yoda{{\sc YODA}\xspace}
\def\delphes{{\sc Delphes3}\xspace}
\def\fastjet{{\sc fastjet}\xspace}
\def\lhapdf{{\sc lhapdf6}\xspace}
\newcommand{\pythia}{{\sc Pythia~8}\xspace}
\newcommand{\madanalysis}{{\sc Mad\-A\-na\-ly\-sis~5}\xspace}

\def\GAMBIT{{\sc GAMBIT}\xspace}
\def\BSMArt{{\sc BSMArt}\xspace}

\def\cpp{{\tt c++}\xspace}

\def\ov{\overline}

\definecolor{maroon}{cmyk}{0, 0.87, 0.68, 0.32}
\definecolor{halfgray}{gray}{0.55}
\definecolor{slha_frame}{RGB}{207, 207, 207}
\definecolor{slha_bg}{RGB}{247, 247, 247}
\definecolor{slha_red}{RGB}{186, 33, 33}
\definecolor{slha_green}{RGB}{0, 128, 0}
\definecolor{slha_cyan}{RGB}{64, 128, 128}
\definecolor{slha_purple}{RGB}{170, 34, 255}

\definecolor{mathematica_frame}{RGB}{207, 207, 207}
\definecolor{mathematica_bg}{RGB}{247, 247, 247}
\definecolor{mathematica_red}{RGB}{186, 33, 33}
\definecolor{mathematica_green}{RGB}{0, 128, 0}
\definecolor{mathematica_cyan}{RGB}{64, 128, 128}
\definecolor{mathematica_purple}{RGB}{170, 34, 255}

\lstnewenvironment{MIN}[1][]{%
  \renewcommand{\thelstnumber}{In[\arabic{lstnumber}]}
  \lstset{language=MathIn,numbers=left,basicstyle=\ttfamily,#1}%
}{%
}

\lstnewenvironment{MOUT}[1][]{%
  \renewcommand{\thelstnumber}{Out[\arabic{lstnumber}]}
  \lstset{language=MathOut,numbers=left,basicstyle=\ttfamily,#1}%
}{%
}

\usepackage{listings}
\lstdefinelanguage{SLHA}{
    morekeywords={block,Block,BLOCK,decay,Decay,DECAY},%
    sensitive=true,%
    morecomment=[l]\#,%
    morestring=[b]',%
    morestring=[b]",%
    morestring=[s]{'''}{'''},
    morestring=[s]{"""}{"""},
    morestring=[s]{r'}{'},
    morestring=[s]{r"}{"},%
    morestring=[s]{r'''}{'''},%
    morestring=[s]{r"""}{"""},%
    morestring=[s]{u'}{'},
    morestring=[s]{u"}{"},%
    morestring=[s]{u'''}{'''},%
    morestring=[s]{u"""}{"""},%
    identifierstyle=\color{black}\ttfamily,
    commentstyle=\color{slha_cyan}\ttfamily,
    stringstyle=\color{slha_red}\ttfamily,
    keepspaces=true,
    showspaces=false,
    showstringspaces=false,
    rulecolor=\color{slha_frame},
    frame=single,
    frameround={t}{t}{t}{t},
    framexleftmargin=6mm,
    numbers=left,
    numberstyle=\tiny\color{halfgray},
    backgroundcolor=\color{slha_bg},
    basicstyle=\footnotesize,
    keywordstyle=\color{slha_green}\ttfamily,
    aboveskip=1.2em,
    belowskip=1.2em,
}

\lstdefinelanguage{HackAnalysis}{
    morekeywords={analyses,settings,Codes,Setup,Variables,new,particles},%
    sensitive=true,%
    morecomment=[l]\#,%
    morestring=[b]',%
    morestring=[b]",%
    morestring=[s]{'''}{'''},
    morestring=[s]{"""}{"""},
    morestring=[s]{r'}{'},
    morestring=[s]{r"}{"},%
    morestring=[s]{r'''}{'''},%
    morestring=[s]{r"""}{"""},%
    morestring=[s]{u'}{'},
    morestring=[s]{u"}{"},%
    morestring=[s]{u'''}{'''},%
    morestring=[s]{u"""}{"""},%
    identifierstyle=\color{black}\ttfamily,
    commentstyle=\color{slha_cyan}\ttfamily,
    stringstyle=\color{slha_red}\ttfamily,
    keepspaces=true,
    showspaces=false,
    showstringspaces=false,
    rulecolor=\color{slha_frame},
    frame=single,
    frameround={t}{t}{t}{t},
    framexleftmargin=6mm,
    numbers=none,
    backgroundcolor=\color{slha_bg},
    basicstyle=\footnotesize,
    keywordstyle=\color{slha_green}\ttfamily,
    aboveskip=1.2em,
    belowskip=1.2em,
}

\lstdefinelanguage{MathIn}{
    morekeywords={Simplify,Eigenvalues},%
    emph={Start,InitUnitarity,GetScatteringDiagrams,BuildScatteringMatrix,MakeSPheno},%
    emphstyle={\color{mathematica_purple}},
    sensitive=true,%
    morecomment=[l]\%,%
    morestring=[b]',%
    morestring=[b]",%
    morestring=[s]{'''}{'''},
    morestring=[s]{"""}{"""},
    morestring=[s]{r'}{'},
    morestring=[s]{r"}{"},%
    morestring=[s]{r'''}{'''},%
    morestring=[s]{r"""}{"""},%
    morestring=[s]{u'}{'},
    morestring=[s]{u"}{"},%
    morestring=[s]{u'''}{'''},%
    morestring=[s]{u"""}{"""},%
    identifierstyle=\color{black}\ttfamily,
    commentstyle=\color{mathematica_cyan}\ttfamily,
    stringstyle=\color{mathematica_red}\ttfamily,
    keepspaces=true,
    showspaces=false,
    showstringspaces=false,
    rulecolor=\color{mathematica_frame},
    frame=single,
    frameround={t}{t}{t}{t},
    framexleftmargin=10mm,
    numbers=left,
    numberstyle=\tiny\color{halfgray},
    backgroundcolor=\color{mathematica_bg},
    basicstyle=\footnotesize,
    keywordstyle=\color{mathematica_green}\ttfamily,
    aboveskip=1.2em,
    belowskip=1.2em,
}

\lstdefinelanguage{MathOut}{
    morekeywords={Simplify,Eigenvalues},%
    sensitive=true,%
    morecomment=[l]\%,%
    morestring=[b]',%
    morestring=[b]",%
    morestring=[s]{'''}{'''},
    morestring=[s]{"""}{"""},
    morestring=[s]{r'}{'},
    morestring=[s]{r"}{"},%
    morestring=[s]{r'''}{'''},%
    morestring=[s]{r"""}{"""},%
    morestring=[s]{u'}{'},
    morestring=[s]{u"}{"},%
    morestring=[s]{u'''}{'''},%
    morestring=[s]{u"""}{"""},%
    identifierstyle=\color{black}\ttfamily,
    commentstyle=\color{mathematica_cyan}\ttfamily,
    stringstyle=\color{mathematica_red}\ttfamily,
    keepspaces=true,
    showspaces=false,
    showstringspaces=false,
    rulecolor=\color{mathematica_frame},
    frame=single,
    frameround={t}{t}{t}{t},
    framexleftmargin=10mm,
    numbers=left,
    numberstyle=\tiny\color{halfgray},
    backgroundcolor=\color{mathematica_bg},
    basicstyle=\footnotesize,
    keywordstyle=\color{mathematica_green}\ttfamily,
    aboveskip=1.2em,
    belowskip=1.2em,
}

\let\origthelstnumber\thelstnumber
\makeatletter
\newcommand*\Suppressnumber{%
  \lst@AddToHook{OnNewLine}{%
    \let\thelstnumber\relax%
     \advance\c@lstnumber-\@ne\relax%
    }%
}

\newcommand*\Reactivatenumber{%
  \lst@AddToHook{OnNewLine}{%
   \let\thelstnumber\origthelstnumber%
   \advance\c@lstnumber\@ne\relax}%
}

\numberwithin{equation}{section}

\title{HackAnalysis 2: A powerful and hackable recasting tool}

\date{}

\begin{document}

\begin{flushright}
\end{flushright}
\begin{center}

\vspace{1cm}

{\bf \LARGE HackAnalysis 2: A powerful and hackable recasting tool}

\vspace{1cm}

\large{Mark D. Goodsell\footnote{goodsell@lpthe.jussieu.fr} 
 \\[5mm]}

{ \sl Sorbonne Universit\'e, CNRS, Laboratoire de Physique Th\'eorique et Hautes Energies (LPTHE), F-75005 Paris, France. }

\end{center}

\abstract{This is the manual for the version 2 of HackAnalysis, a powerful, lightweight, versatile and, most importantly, hackable, recasting tool. New features in this version include: compressed event format storage for ultra-fast development; integration of new physics and mathematics routines via RestFrames and Eigen; automatic computation of systematic uncertainties; an interface to ONNX for neural networks; easy implementation of new models via QNUMBERS blocks; a python package for interfacing to  pyhf, spey and in-built statistics routines for fast computation of experimental limits; improved integration with BSMArt for fast scanning, and a new batch running/convergence check. Several new (electroweakino) analyses are included with this release. 
}

\vspace{0.2cm}

\newpage
\setcounter{footnote}{0}
\tableofcontents

\section{Introduction}
\label{SEC:INTRO}


The LHC is currently collecting unprecedented amounts of data at the energy frontier, and the experimental collaborations are working on new techniques for extracting as much information as possible from it. Among the vast number of papers already written using that data are hundreds of searches for physics beyond the Standard Model. All of these place exclusion limits on the parameter space of some models, and some reveal excesses that hint at new physics. The task to glean as much as possible from this information then falls on the shoulders of theorists, because there is (currently) no way to look at the data and directly work backwards to nature's laws. Instead, theorists must propose models and then attempt to test them against the data; since the data are not public, it is first necessary to reproduce the conclusions of the experimental papers in a code, hopefully written in as general way as possible so that it can be applied to other models. This task is known as \emph{recasting}. The procedures for doing this have become well-established; see \cite{Abdallah:2020pec} for a review and references therein. However, they are constantly in development, with recent white papers on publishing statistical models \cite{Cranmer:2021urp}, data preservation \cite{Bailey:2022pdq} and machine-learning analyses \cite{Araz:2023mda}. 

An impressive number of LHC \emph{measurements} -- meaning, analyses of Standard Model processes -- have been recast in the tool {\sc RIVET} \cite{Buckley:2010ar,Buckley:2019stt,Bierlich:2019rhm}. These can be used to constrain BSM models too via the deviations induced in SM processes; this can be automatically checked via  {\sc Contur} \cite{Butterworth:2016sqg,Buckley:2021neu}, which is becoming a valuable tool complementary to the ones below, particuarly since {\sc RIVET} contains many innovative features. 

On other other hand, for Beyond the Standard Model (BSM) \emph{searches}, other tools are used. In a category of its own among these is {\tt SModelS} \cite{Kraml:2013mwa,Alguero:2020grj,Ambrogi:2017neo,Ambrogi:2018ujg,Khosa:2020zar,Alguero:2021dig,MahdiAltakach:2023bdn}, which maps the components of a model onto simplified model topologies for which results have been computed as efficiency grids; this allows very fast results and can be accurate provided that the map from the model to the simplified topologies is straightforward. To produce the efficiency grids, and for more general results, it is necessary to use a more complete simulation.

For recasting full simulations, there exist three major public frameworks: \madanalysis \cite{Conte:2012fm,Dumont:2014tja,Conte:2018vmg,Araz:2020lnp,Araz:2021akd}, {\sc ColliderBit} \cite{GAMBIT:2017qxg,GAMBIT:2018gjo,Kvellestad:2022uzz,GAMBIT:2023yih} (part of \GAMBIT) and {\sc CheckMATE} \cite{Drees:2013wra,Dercks:2016npn,Desai:2021jsa} with different motivations. \madanalysis contains interactive tools for analysing event samples and visualising them, with an expert mode and Public Analysis Database (PAD) which provide a framework for analysing samples using {\tt c++} code that is as close to the logic of the experiment as possible, producing a cutflow for each signal region that can be compared to those provided. The PAD therefore contains analyses, many contributed by the experimental collaborations themselves, where the validation has been performed to a high standard, incuding a recasting note for each describing what has been done, what checks were made, etc. This should be regarded as the gold standard in terms of reproducability and transparency. The PAD can be (and is regularly) used for providing constraints on model parameter spaces.

On the other hand, the aim of {\sc CheckMATE} is to check whether a given parameter point is excluded or not, abstracting the inner workings as much as possible; while the aim of {\sc ColliderBit} is to map out the likelihood of a wide range of parameter space as quickly as possible. Hence {\sc ColliderBit} is optimised for speed, so makes some compromises in terms of the simulation and reconstruction. 

However, as new analysis techniques are conceived and put into practice by our experimental colleagues, it is often a challenge to replicate or approximate them in monolithic recasting packages, precisely because they break some assumption in the workflow, from adding new ways of reconstructing objects such as jets, to adding neural networks, etc. For the same reason, the major recasting packages are rarely used for proposing new analysis techniques (except by their authors). It is therefore useful to have a flexible tool that can be hacked by the user where new features can be added without breaking fundamental assumptions. That was the original aim of \hackanalysis: it was not possible/straightforward to accommodate necessary aspects of long-lived particle searches in the major recasting frameworks at the time of its inception and original release documented in \cite{Goodsell:2021iwc}.

The goals of \hackanalysis are to have a tool that can be: maximally flexible; with a minimum of external dependencies; as fast as possible without sacrificing accuracy of simulation; and to reduce as much as possible the repetitive and tedious aspects that can be found when implementing and running new analyses. Such analyses should then be translated to the major platforms for use in standard workflows or global fits; it is not the intention to build a competitor database of analyses. 

The original version of \hackanalysis took advantage of the move towards fast detector simulation based on unfolding as used by default in {\sc RIVET} and \GAMBIT and later added to \madanalysis via the SFS module \cite{Araz:2020lnp}. In line with other platforms except for {\sc CheckMATE} it avoids the {\tt root} software used by the experimental collaborations for handling four-vectors and plotting because it is difficult to install and has large overheads. Rather than creating another implementation, the {\tt heputils}\footnote{\url{https://gitlab.com/hepcedar/heputils}} header files used by \GAMBIT are used to handle basic objects; however, over time it has been necessary to enhance them with many extra features, and the namespace has been renamed to {\tt HEP}. Due to the same or similar basic functions, however, this should aid porting of analyses from \hackanalysis to \GAMBIT. In line with {\sc RIVET}, {\sc YODA} \cite{Buckley:2023xqh} is used for histogramming and plotting. On the other hand, the \emph{analysis} structure was deliberately kept as close as possible to that of \madanalysis: signal regions are defined with very similar commands and cuts are applied in a similar way. The main novel features of the original code were that the ``detector'' simulation had a finite size and took into account metastable particles; and the ability to accept different types of inputs according to the preferences or speed/accuracy requirements of the user (so that, e.g. when debugging, events could be analysed ultra-rapidly and then more realistic -- yet as fast as possible -- simulations could be used when fine-tuning).



Since the original release, \hackanalysis has been used to develop new analyses with novel features, mentioned in subsequent papers\cite{Araz:2021akd,Agin:2023yoq,Agin:2024yfs}. This paper collects and documents these features, as well as some new ones, and is accompanied by the release of version {\tt 2.1} of the code. The new features include:
\begin{itemize}
\item Simple inclusion of new particles, described in section \ref{SEC:QNUMBERS}.
  \item Possibility of having multiple (user-provided) detector simulations for the same set of eventss, described in section \ref{SEC:DETECTORS}.
  \item A compressed event format for ultra-rapid analysis development, described in section \ref{SEC:HAEVENT}.
    \item Automatic systematic uncertainties in cutflows, described in section \ref{SEC:SYSTEMATICS}. 
\item Integration with \BSMArt \cite{Goodsell:2023iac,Faraggi:2023jzm} for fast scanning, and for operating parallel \MadGraph \cite{Alwall:2014hca} gridpacks to process batches of events. This also allows for convergence checking, ensuring that only \emph{just enough} events are simulated. This is described in section \ref{SEC:BSMART}.
\item Inclusion of the {\tt restframes} package, the MT2 variable and Mathematics routines via the {\tt Eigen} package: see section \ref{SEC:RESTFRAMES}.
\item Linking to ONNX for machine-learning analyses, described in section \ref{SEC:ONNX}.
\item {\tt json} file output of cutflows and efficiencies.
\item Scripts for merging cutflows and creating tex output.  
\item Several new analyses, summarised in section \ref{SEC:ANALYSES}. 
\item New statistical routines for computing exclusion limits and significances using {\tt pyhf}, {\tt spey} \cite{Araz:2023bwx} and toy-based/asymptotic calculators. These are described in section \ref{SEC:STATS}, which includes a novel treatment of the significance of excesses in the CMS monojet search.
\end{itemize}
Instructions on implementing new analyses are given in section \ref{SEC:YOURANALYSIS}.

\section{Installation}
\label{SEC:Installation}

The code can be downloaded from \url{https://goodsell.pages.in2p3.fr/hackanalysis/}. While effort has been expended to keep the number of external packages as small as possible, \HackAnalysis relies on \hepmcv \cite{Dobbs:2001ck}\footnote{https://hepmc.web.cern.ch/hepmc/releases/hepmc2.06.09.tgz}, {\tt fastjet}\cite{Cacciari:2011ma}\footnote{\url{https://fastjet.fr}}, \yoda \cite{Buckley:2023xqh}\footnote{\url{https://yoda.hepforge.org/}} and \pythia \cite{Sjostrand:2014zea,Bierlich:2022pfr}\footnote{\url{https://pythia.org}} as external packages, and these must all be built first. In particular, \hepmcv should be built with the units of GeV and mm. The paths to the libraries must then be specified in the file {\tt Paths.inc} in the \hackanalysis installation directory:
\begin{lstlisting}[language=SLHA,title=Paths.inc]
ZLIBpath=
###### The following paths need to be specified:

hepmcpath=
pythia8path=
fastjetpath=

YODApath=
#### Here a subtlety, need e.g.:
#### YAMLpath=<where your YODA code is>/YODA-1.8.5/src/yamlcpp
YAMLpath=

### If you have installed ONNX, uncomment this and give the path to
### enable ONNX analyses
#ONNXPATH=
###

USE_RESTFRAMES=1
USEYODATWO=1 
\end{lstlisting}
It is not essential to specify a path for ZLIB, but omitting it will disable compression. If the flag {\tt USE\_RESTFRAMES} is removed, then the relevant code will not be included and the analyses that use it will be ignored. The flag {\tt USEYODATWO} should be removed or set to undefined if using an \emph{older} version of \yoda; as of version 2 the include paths changed in \yoda.

Fortunately, to ease all of these tasks, \emph{all} of the relevant packages (including \hackanalysis itself) can be installed via a provided script, which will also edit the {\tt Paths.inc} file. Everything can then be installed in just three lines:
\begin{lstlisting}[]
  ./installColliderTools.py --All
  cd HackAnalysis
  make
\end{lstlisting}
In principle the build command could be rolled into the script, but it can be (much) faster to call {\tt make} with the option {\tt -j4}, or {\tt -j8} etc to use more cores. The above will also install \madgraph and {\tt lhapdf6} at the same time. It will also produce scripts called {\tt SETUP\_COLLIDERS.sh} and {\tt configure\_colliders.csh} for {\tt bash} and {\tt csh/tcsh} respectively, which when called via {\tt source} (e.g. {\tt source SETUP\_COLLIDERS.sh}) will configure the library and {\tt python} paths for all installed packages -- this is particularly important for systematics in \madgraph and for producing plots with \yoda.

\section{Running \hackanalysis}
\label{SEC:RUNNING}

Once built, the code will produce four executables:
\begin{itemize}
\item {\tt analysePYTHIA.exe} generates events and showers them internally using \pythia.
\item {\tt analysePYTHIA\_LHE.exe} reads {\tt .lhe} or {\tt .lhe.gz} files of hard events, and showers the events internally using \pythia. 
\item {\tt analyseHEPMC.exe} reads \hepmc files of showered events. 
\item {\tt analyseHAEVENT.exe} reads custom event files that have already been processed by the detector simulation within \hackanalysis and stored in an ultra-compact format: see section \ref{SEC:HAEVENT}.
\end{itemize}
Each of the above takes as input a {\tt yaml} file which contains the necessary settings for a given run.  Examples are given in the directory {\tt InputFiles}. The format of the file is the same for all four executables, but some of the settings are only relevant for one or two. A complete list of the possible settings is:
\begin{lstlisting}[language=HackAnalysis]
---
analyses:
 - <analysis name 1>
 - <analysis name 2>
 - <etc>
settings:
  nevents: <number of events to simulate>
  cores: <number of cores to run> # ignored by analyseHEPMC.exe
  Include Pileup: <true> or <false>
  Config file: <file containing pythia commands> # only for analysePYTHIA.exe and analysePYTHIA_LHE.exe
  LHE file: <LHE or ha file>    # only for analysePYTHIA_LHE.exe and analyseHAEVENT.exe
  HEPMC file: <HEPMC filename> # only for analyseHEPMC.exe
  Efficiency Filename: <filename for SLHA-style efficiency list>  
  Cutflow Filename: <filename for cutflows>  
  Histogram Filename: <filename for yoda histograms> 
  JSON file: <json filename for outputs>  l
  Cross-section: <optional, float>  # if not specified, will use value from LHE or pythia
  Store Events: <true> or <false> 
  Store Hadrons: <true> or <false>
  QNUMBERS file: <SLHA file containing QNUMBERS of new particles beyond the MSSM>
new particles:
  <pdg>:
     3Q: <3 times electric charge>
     colour: <dimenson of colour rep>
\end{lstlisting}
The {\tt Efficiency Filename} corresponds to an {\tt SLHA}-style output file containing the efficiencies for each analysis and region; it can be read by {\tt zslha.py} in \BSMArt and looks like, for example:
\begin{lstlisting}[language=SLHA,title=Efficiency File]
BLOCK PROCESS ATLAS_SUSY_2018_16
XS   0.000334093   # (fb) 
Events  1564329
BLOCK EFFICIENCIES ATLAS_SUSY_2018_16 # name, eff, rel. uncert.
SR_E_lT_eMLLa   0.000000e+00   1.000000e+00
SR_E_lT_eMLLb   4.320019e-05   1.216421e-01
SR_E_lT_eMLLc   5.072136e-05   1.122612e-01
...
\end{lstlisting}

The cutflows are all given in one human-readable file with information about the region, number of events, sum of weights, and uncertainties, e.g. in this example with leading-order generation for metastable chargino production passing through analyses {\tt DT\_CMS} and {\tt HSCP\_ATLAS}:
\begin{lstlisting}[language=SLHA,title=Cutflow File]
============================= SR1_2017 =============================
Initial Events : 20007, 186.541 -> efficiency: 1 +/- 0%
MET : 3660, 34.125 -> efficiency: 0.182936 +/- 1.49413%
OneGoodJet : 2229, 20.7827 -> efficiency: 0.111411 +/- 1.99662%
JetAngleCuts : 2037, 18.9925 -> efficiency: 0.101814 +/- 2.09985% 
...
4 Layers 2017 : 3, 0.0279713 -> efficiency: 0.000149948 +/- 57.7307%
...
============================= CAND1_0 =============================
Initial Events : 20007, 186.541 -> efficiency: 1 +/- 0%
Trigger : 15052, 140.342 -> efficiency: 0.752337 +/- 0.405633%
Preselection : 11486, 107.093 -> efficiency: 0.574099 +/- 0.608934%
1Tight : 3618, 33.7334 -> efficiency: 0.180837 +/- 1.5047%
mTOF>175 : 3384, 31.5517 -> efficiency: 0.169141 +/- 1.56693%
...
\end{lstlisting}
If systematic error information is available (see section \ref{SEC:SYSTEMATICS}) then it will be printed in this file too. 

% ...

The new output json file contains all information about every cutflow, including sums of weights, sums of squares of weights, systematics information etc, in a format that can easily be read into {\tt python} scripts for processing: there are scripts provided for producing \LaTeX\  cutflows, performing statistics checks etc that are described below. 

\subsection{Generating events with \pythia}

For common Standard Model processes, for the MSSM and for simple BSM models such as a $Z'$ boson, \pythia is capable of generating hard events involving two-body final states, because it has the matrix elements hard-coded. \GAMBIT uses this functionality by default, and provides GUM \cite{Bloor:2021gtp} to do the same for other models: \madgraph can generate matrix element code that can be used by \pythia. The generation of events in this way is \emph{much} faster than the alternatives, especially since no intermediate files need be written to disk, so it can be used as a first approximation for available models (or indeed it is what the experimental collaborations use for certain exotic higher-dimensional models). On the other hand, the computation of missing energy and modelling of initial/final state radiation is not accurate when relying only on \pythia; for monojet searches in particular the results cannot be relied upon, where additional initial/final state radiation should be included in the matrix element, and one of the other approaches should be used. Hence this option is recommended for quick testing or exotic models only. 

As an example yaml file for invoking \hackanalysis to generate events via \pythia is included with the package:
\begin{lstlisting}[language=HackAnalysis,title=LOpythia.yaml]
---
analyses:
 - DT_CMS
 - HSCP_ATLAS
 
settings:
  nevents: 1000
  cores: 1
  Include Pileup: false
  Efficiency Filename: LO.eff
  Cutflow Filename: LO_cf.eff
  Histogram Filename: LO.yoda
  Config file: LO.cfg
\end{lstlisting}
This should be called via
\begin{lstlisting}[]
./analysePYTHIA.exe LOpythia.yaml
\end{lstlisting}
Here the important lines are the number of events ({\tt nevents}) and the option {\tt Config file}, which specifies the extra commands passed to \pythia. In this example, the file {\tt LO.cfg} contains:
\begin{lstlisting}[language=HackAnalysis,title=LO.cfg]
Beams:eCM = 13000
Beams:idA = 2212
Beams:idB = 2212
SLHA:file = AMSB_chargino_700GeV_1000cm_Isajet780.slha
SigmaDiffractive:mode = 2
SUSY:qqbar2chi+chi- = on
SUSY:qqbar2chi+-chi0 = on
\end{lstlisting}
which means that electroweakino events are generated at centre of mass energy of 13 TeV with the specified {\tt SLHA} parameter card; this example was taken from those provided by CMS for the disappearing track analysis \cite{CMS:2018rea,CMS:2020atg}.

The running of \hackanalysis with event generation in \pythia can be handled automatically by \BSMArt using the tool {\tt HackAnalysis\_LO}, as described in the \BSMArt manual \cite{Goodsell:2023iac}. 

\subsection{Generating events with \madgraph and showering with \pythia}

This option for running \hackanalysis involves specifying an entry for the {\tt LHE file} in the yaml input. If running in single core mode, the entire filename for the {\tt les houches event} file should be given here. If, on the other hand, the code is to be run in multicore mode, then the events should be split into several files with suffixes {\tt \_0.lhe.gz}, {\tt \_1.lhe.gz}, ... with one file per core. In order to specify the name of this file, the entry {\tt LHE file} should only contain the prefix. The default choice is for the files to be placed in a subdirectory called {\tt Split} with files {\tt Split/split\_0.lhe.gz}, {\tt Split/split\_1.lhe.gz}, ..., and in which case the yaml file should have the line:
\begin{lstlisting}[language=HackAnalysis]
settings:
  LHE file: Split/split
\end{lstlisting}
It is not necessary for the files to be compressed (if {\tt GZIP} is not available, for example): the code will also look for files {\tt Split/split\_0.lhe} etc and work with them.

Splitting the files in this way from one generated by the user's favourite event generator is facilitated by a script {\tt splitlhe.py} provided with the code. Alternatively, the generation of event files and running of \hackanalysis can be handled by \BSMArt using the tool {\tt MadGraphHackAnalysis}: the original instructions can be found in the \BSMArt manual \cite{Goodsell:2023iac}, and new instruction for operating in an ultra-fast gridpack mode involving batches of events and convergence checking is described in section \ref{SEC:BSMART} below.

The other essential file for this case is the configuration file for \pythia. In particular, we require the settings for activating jet merging by a chosen scheme. For example, the included example file for MLM merging with up to two jets contains: 
\begin{lstlisting}[language=HackAnalysis,title=MLM.cfg]
Init:showChangedSettings=off
Init:showChangedParticleData = off  ! not useful info here
Next:numberShowInfo       = 0    ! avoid unnecessary info
Next:numberShowEvent      = 0    ! avoid lengthy event listings
Next:numberCount=0

Beams:setProductionScalesFromLHEF=on

JetMatching:merge            = on
JetMatching:scheme           = 1
JetMatching:setMad           = off
JetMatching:qCut             = 175.0

JetMatching:coneRadius       = 1.0
JetMatching:etaJetMax        = 10.0
JetMatching:nJetMax          = 2

JetMatching:nQmatch = 4
JetMatching:clFact = 1.0
\end{lstlisting}
The \BSMArt manual \cite{Goodsell:2023iac} contains further instructions on how the {\tt qCut} parameter can be set dynamically during a scan; other example files provided (and also those with the \pythia installation) show how other merging schemes can be used.

\subsection{Reading \hepmc files}

Reading \hepmc files is perhaps the simplest way of operating \hackanalysis. Once a \hepmc file has been generated, the filename need only be specified in the yaml file as {\tt HEPMC file} and the appropriate executable called, or the path can be specified on the command line:
\begin{lstlisting}[]
./analyseHEPMC.exe HEPMC.yaml [Optional: hepmc filename]
\end{lstlisting}
It is advisable in this mode of operation to specify a cross-section in the yaml file, because that information is not stored in the \hepmcv file. This option is equivalent to running \madanalysis because the reader will operate using only a single core, and it is therefore slower than other methods. However, it is probably the only one availble for handling NLO output. 

The generation of \hepmc files and running of \hackanalysis can be handled automatically through \BSMArt via the tool {\tt MadGraphHackAnalysis}, as detailed in the \BSMArt manual \cite{Goodsell:2023iac}. 

\subsection{Reading \hackanalysis events}

This new feature allows pre-generated events to be reprocessed ultra-rapidly by \hackanalysis. If the option {\tt Store\ Events} is set to {\tt true} in the yaml file for one of the other analysis codes, then events will be written to disk, one per core, with names {\tt hackanalysis\_events\_0.ha.gz}, {\tt hackanalysis\_events\_1.ha.gz}, ... These can then be read by the {\tt analyseHAEVENT.exe} executable. Similarly (and perhaps confusingly) to {\tt analysePYTHIA\_LHE.exe},  the parameter used is again {\tt LHE\ file}, which should be specified for multicore mode as:
\begin{lstlisting}[language=HackAnalysis]
settings:
  LHE file: hackanalysis_events
\end{lstlisting}
(or the full filename should be used for single core mode). The analysis can then be run:
\begin{lstlisting}[]
./analyseHAEVENT.exe HAEVENT.yaml
\end{lstlisting}
Run times of order seconds are to be expected even for very large numbers of events: more details about the event format are given in section \ref{SEC:HAEVENT} below.

\section{Inclusion of models with new particles}
\label{SEC:QNUMBERS}

Including new particles and interactions in monte-carlo tools is now standard via interfaces such as {\tt UFO} \cite{Degrande:2011ua,Darme:2023jdn}. In parton showers and in {\tt SModelS} however we do not need information about particle interactions beyond their charge and QCD quantum numbers, and this can be supplied via {\tt QNUMBERS} blocks in an {\tt SLHA} file. Hence in \HackAnalysis it is now possible to include new particles by giving the name of a file containing {\tt QNUMBERS} blocks, via the parameter {\tt QNUMBERS file} in the input YAML file. 

Alternatively the quantum numbers can by input directly via a block labelled {\tt new particles} where each new particle is specified by its pdg id, and has entries {\tt 3Q} for three times the charge and {\tt colour} for the QCD rep, e.g.:
\begin{lstlisting}[language=HackAnalysis]
---
analyses:
 ...
settings:
  QNUMBERS file: QNUMBERS_MSSM.slha
new particles:
  5000039:
     3Q: 0
     colour: 1
\end{lstlisting}

New neutral, colourless particles will be identified by the code as invisible (with the obvious caveat that composite states cannot be handled in this way) whereas colourful states will be treated as hadronic for the purpose of the detector simulation. 

\section{Multiple ``detector'' simulations}
\label{SEC:DETECTORS}

The full chain of simultation as performed by the experimental collaborations is: (1) hard event generation; (2) showering and hadronisation; (3) detector simulation; (4) jet clustering and object reconstruction, including isolation requirements; (5) compensation for detector effects; (6) event analysis. Since steps (3), (4) and (5) are always closed source with only sparse details released to the public, in recasting we must always attempt to make approximations.

The standard approach is to use a detector simulation -- usually \delphes \cite{deFavereau:2013fsa} -- which also invokes \fastjet to perform jet clustering, and which will perform object reconstruction and some rudimentary isolation checks. However, this is usually handled by passing the enormous \hepmc files for the event sample, and returns a {\tt root} file, meaning that the analysis code becomes dependent on {\tt root}. Moreover, there is actually some information lost in the process, because \delphes can never implement the compensation effects applied by the experiments -- they are simply unknown. The use of machine learning algorithms for object reconstruction, jet vertex tagging etc have become widespread during Run 2 at the LHC, and these are all private codes. The goal of all these private compensatory and sophisticated identification steps is to make the detectors as close to reconstructing particle-level information as possible, and it is impressive how much progress has been made. These can make a simplified perfect detector actually perform poorer than the real thing.

The other approach to steps 3 to 5 above, taken by {\sc RIVET}, \GAMBIT and the SFS module of \madanalysis \cite{Araz:2020lnp}, is to use particle-level information instead, and reconstruct jets from hadrons. Detector effects are then encapsulated in efficiencies and smearing of momenta, which are performed stochastically and require information from the experiments. This has many advantages: it is (much) faster; does not require handling large files (or to use {\tt root}); it is more independent of the algorithms and techniques used by the experiment; it is often easier (if not actually easy) to find information on uncertainties and efficiencies from the experiments; it allows for more flexibility. This is also the approach taken in \hackanalysis.

To implement a particle-level reconstruction, however, there are different approaches possible. In \madanalysis SFS and \GAMBIT the jet clustering and object reconstruction are performed in one step. In the second step, efficiencies and smearing are applied. In \madanalysis SFS a selection of ``detector cards'' analagous to \delphes cards are used, and several analyses may use one card. In \GAMBIT efficiencies and smearing functions are provided for both ATLAS and CMS, and these are applied to the groups of reconstructed objects in each analysis. However, at the level of the event reconstruction, the detector is assumed to be infinitely large in both cases, so that all unstable particles are decayed and only the final products are kept; the main difference between the two is that in \GAMBIT particles that come from decays of light mesons and baryons are treated as part of jets (removing leptons from such decays is one of the purposes of isolation routines) whereas \madanalysis SFS allows the user to demand the computation of isolation cones around each particle or track so that object cleaning/isolation can be performed within the analysis.

In both of the above approaches, it is difficult to take into account metastable or long-lived particles. In order to implement these in \madanalysis, tracks representing the particles were added; but the computation of missing transverse energy must be corrected inside analyses to take into account for the finite size of real detectors. On the other hand, this approach is well-suited to running large number of different analyses over the same sample, since the jet clustering is done only once. Since \hackanalysis is intended to be as flexible and fast as possible for developing new analyses and adding new features, instead new detectors are implemented as separate event-filling routines, which can be written as required. This means that the jet clustering and object selection is done for each ``detector;'' this will be somewhat slower if many detectors are required (compensated with respect to \madanalysis by the possibility of using multiple cores), but the prize is much greater flexibility.

As an example, the jet cluster radius is actually specified within the filling routine, so to implement Fat Jets there is a fill function called {\tt FATJET}; this is obviously inefficient if many different types of jets are required for different analyses, but when working on new analyses or recasts it is very helpful: for example, it is straighforward to include advanced \fastjet features such as {\tt GridMedianBackgroundEstimator} (and some commented-out code is included which does exactly that in the {\tt include/Detectors/Default\_Pythia.h} file) -- this would be extremely laborious to do in other approaches. 

A pre-defined detector can be selected within an analysis by defining the class member ``DetectorFunction,'' e.g.:
\begin{lstlisting}[language=C++,title=ATLAS\_EXOT\_2018\_06.cpp]
void ATLAS_EXOT_2018_06::init() {

this->DetectorFunction="ATLAS";
\end{lstlisting}
Included detectors as of version 2.1 are {\tt DEFAULT, ATLAS, FATJET}. In order to define a new detector routine, the code for it should be added to an existing header file or a new one within the {\tt Detectors} subdirectory, and an entry in a map from the function names to the relevant functions added to the files {\tt fillevent.h} and {\tt filleventP.h}, which handle the ``filling'' of events when reading \hepmcv files and when running \pythia respectively. The included files contain the lines:
\begin{lstlisting}[language=C++,title=fillevent.h]
  std::map<std::string, std::function<void(HEP::Event &OutEvt, HepMC::GenEvent* evt, int npileups, std::vector<HEP::PileupEvent*> &pileups, std::mt19937 &engine)>> hepmcFillMap =
{ 
  {"DEFAULT", filleventHEPMC},
  {"ATLAS",filleventHEPMC_ATLAS},
  {"FATJET",filleventHEPMC_FatJet}
};
\end{lstlisting}
and
\begin{lstlisting}[language=C++,title=filleventP.h]
std::map<std::string, std::function<void(HEP::Event&, Pythia8::Event&, double, int, std::vector<HEP::PileupEvent*>&, std::mt19937&)>> pythiaFillMap =
{ 
  {"DEFAULT", filleventPYTHIA}, 
  {"ATLAS", filleventPYTHIA_ATLAS},
  {"FATJET",filleventPYTHIA_FatJet}
};
\end{lstlisting}

The main differences between the {\tt ATLAS} and {\tt DEFAULT} functions is that the former takes the \GAMBIT approach to including the decay products of light hadrons only in jets and not as separate analysis objects (i.e. we do not treat leptons from pion decays as signal leptons); and also the geometry of the {\tt DEFAULT} detector correponds to the physical dimensions of CMS compared to ATLAS for the other. Since the original reason for developing \hackanalysis was to implement the recast of the CMS disappearing track search, the finite size of the detector \emph{is} automatically taken into account in the missing energy computation and metastable charged particles are defined as analysis objects.

In all of the included detectors, no efficiencies are implemented, and so most/all \GAMBIT or \madanalysis SFS analyses could be implemented using efficiency/smearing functions applied within the analyses themselves (this also means that an analysis can be quickly developed in \hackanalysis and then easily ported to another format). To this end, many convenience functions are included, which can be seen in the files {\tt include/isolation.h} and {\tt include/smearing.h}.

\section{Compressed event format for ultra-rapid analysis development}
\label{SEC:HAEVENT}

In the procress of developing an analysis it is often necessary to run the analysis code over a test sample multiple times. This can very time consuming if that sample is large, even if it is in the form of a {\tt root} file produced by {\tt Delphes}, and even more so for fast collider simulations processing a \hepmc file. To avoid performing the same task repeatedly, the {\tt LHE} event format was extended to allow the storing of reconstructed objects (instead of just hard-scattering events) where reconstructed jets would be represented via the {\tt pdg} code $21$ (for gluons) and b-tagged jets via the code $5$ (for b-quarks). It was also proposed that any other desired objects could be stored via other XML tags.

While the \emph{output} of such {\tt lhe} files is possible in \madanalysis, reading them in requires rewriting the analysis to accept the data from the  ``monte-carlo'' objects: there is currently no straightforward way to use these files instead of a \hepmc or {\tt root} file. This is partly deliberate, because of the loss of information about isolation: only final states are stored. Another drawback is that, in {\tt lhe} files, the production vertex for particles are not stored: while lifetimes can be specified, so decaying particles can be stored, non-prompt particles that originate from the decay of a metastable mother simply cannot be accommodated. Hence it is necessary to modify the format further.

In \hackanalysis, it is now possible to store reconstructed objects from the detector simulation stage in a similar format where the objects are grouped by XML tags. Setting {\tt Store Events: true} in the {\tt yaml} input file will cause the events from the first detector simulation that is run to be stored; if multiple cores are being run then these will be stored in separate (compressed) files, named {\tt hackanalysis\_events\_0.ha.gz}, {\tt hackanalysis\_events\_1.ha.gz}, ... etc, with as many files as cores are used. A typical event might appear as (where I have deleted some lines of jets for reasons of space):
\begin{lstlisting}[language=SLHA]
<Event>
<weights>
3.84643 
</weights>
<ptmiss>
113.17 -60.3375 359.526 735.925 
</ptmiss>
<stable particles>
1 -13 3 0.350215 -0.205733 0.828542 0.928775 0 0 0 0 0 
</stable particles>
<invisible particles>
2 14 0 0.014767 1.13728 4.54119 4.68145 0 0 0 0 0 
3 1000022 0 162.93 12.2614 -100.503 243.51 0 0 0 0 0 
4 1000022 0 -49.7751 -73.7362 455.487 487.733 0 0 0 0 0 
</invisible particles>
<HSCPs>
</HSCPs>
<Jets>
-57.8146 24.9538 -284.179 291.134 0 
-32.2594 5.7379 -67.4445 75.3996 0 
-11.3639 13.2398 48.61 52.0372 0 
-4.75325 3.42486 4.92632 7.91603 0 
-2.10755 3.61075 -7.9571 9.04343 0  
</Jets>
\end{lstlisting}
For stable and metastable particles the entries are: a unique identifier, the {\tt pdg} number, the charge, the momenta, production vertex coordonates and either the decay vertex (if the particle decays) or zero (if it does not) are stored. For jets the information are the four components of their momenta and one or zero for the b-tag. 

If {\tt Store Hadrons: true} is set in the {\tt yaml} file, then the hadronic components of the jets are all stored, for the purpose of allowing the user to compute their own isolation checks. However, since this has the effect of greatly increasing the size of the files and slowing down the writing and reading, the default is to drop them. This is similar to the approach of {\tt Delphes} which performs rudimentary isolation checks (if desired) and stores the ``towers'' of hadronic activity if instructed so that the user can impose more sophisticated isolation requirements (as often done by the experimental collaborations) at the expense of larger files. In comparison, {\tt GAMBIT} only impose isolation as an efficiency, which is of course much faster: this can also be done (in effectively exactly the same way) in \HackAnalysis and, if possible, is the favoured approach to take, since it speeds up the analysis development and running.

As an example, for $100k$ simulated $p p \rightarrow \tilde{\chi}_2^0 \tilde{\chi}_1^\pm, \tilde{\chi}_2^0$ hard events, the {\tt lhe.gz} files take up roughly $19$MB;  after parton showering, MLM merging and fast detector simulation we have about $50k$ events, and without storing the jet constituents the {\tt .ha.gz} take roughly $10$MB of disk space. Contrast this to the size of the \hepmc file of $7.1$GB! Exchanging such files via the disk and processing them in single cores is clearly extremely inefficient; to a certain extent, this is why \hackanalysis exists. Since one {\tt .ha.gz} file is created per core, the reading of these events is automatically performed using multiple cores, too, with the result that the reprocessing through an analysis of even very large number of events takes time of order a second, rather than several minutes (or more): this is a major boon when tweaking new recasts, where frequent small changes need to be compared.

\section{Automatic systematic uncertainties in cutflows}
\label{SEC:SYSTEMATICS}

The information about systematic uncertainties due to the variations of scales and pdf sets are automatically included in {\tt les houches event} files by \madgraph when \lhapdf is available: the information is stored in groups of weights. \pythia reads this weight information but contains no built-in functions for making use of them. As a new feature in \hackanalysis, when using {\tt analysePYTHIA\_LHE.exe} (i.e. when reading in {\tt les houches event} files and showering with \pythia internally) this information about the different weight groups is stored and retained through the computation of cutflows. There are different algorithms specified for how to use the weights: typically these are ``envelope,'' but the pdf sets have different algorithms such as ``hessian'' etc, and the way to compute the uncertainty is then different for different pdf set: it is intended that \lhapdf is used to compute pdf set uncertainties. Rather than implementing all the different algorithms, and since it is not straightforward to use \lhapdf for this purpose, in version 2.1 of \hackanalysis only the uncertainties due to groups specifying the envelope method are used, and then the uncertainty for separate weight groups is added in quadrature. Since the pdf uncertainty is often smaller than the other sources, this still gives a reasonable estimate of the systematic uncertainty on each cut.

The systematics are computed automatically whenever the information is present in the {\tt les houches event} files, and output in the efficiency file, cutflow file and json output. \LaTeX\ cutflows can then be produced automatically from the json output file including this information using the script {\tt ha\_texoutput.py} in the {\tt Statistics} folder: examples of such cutflows are given in tables \ref{TAB:LM}, \ref{TAB:MM}, \ref{TAB:HM}. These are cutflows for the analysis ATLAS-SUSY-2019-08, with signal process $p p \rightarrow \tilde{\chi}_2^0 \tilde{\chi}_1^\pm$ plus up to two additional jets production where the decays are $\tilde{\chi}_2^0 \rightarrow \tilde{\chi}_1^0 + h, h \rightarrow b \ov{b}$ and $\tilde{\chi}_1^\pm \rightarrow \tilde{\chi}_1^0 + W^\pm, W^\pm \rightarrow \ell + \nu$. The $\tilde{\chi}_2^0/\tilde{\chi}_1^\pm$ states are assumed to have identical masses, heavier than the neutralino.

The reader will note that the statistical uncertainty of the initial number of events is zero (as it must be by definition) but may be confused by the fact that the statistical uncertainty is not vanishing: for just the first cut, the statistical uncertainty is given as the proportional uncertainty on the total of the weights. This is roughly the uncertainty on the cross-section. 


\section{Integration with \BSMArt for fast scanning and convergence checking}
\label{SEC:BSMART}

While \GAMBIT perfectly integrates scanning algorithms with the codes run, there is no canonical method for running other analysis tools over a set of points. \BSMArt \cite{Goodsell:2023iac,Faraggi:2023jzm} was however created to implement both simple and sophisticated scanning algorithms and pilot a range of tools in the \SARAH \cite{Staub:2013tta,Goodsell:2017pdq,Braathen:2017izn,Goodsell:2018tti,Goodsell:2020rfu,Benakli:2022gjn} family. With the latest versions of \BSMArt the operation of \hackanalysis in LHE mode (the fastest method for generating realistic events for the LHC) has been greatly enhanced with two new features: (1) the operation of gridpacks and (2) convergence checking.

If the option {\tt GridPack} is set to {\tt True} in the input json file, then for each parameter point tested, a \madgraph gridpack will be generated. A gridpack is able to generate a required number of events desired and runs on a single core. When invoked in read-only mode, several gridpacks can be run in parallel, and this removes bottlenecks in the event generation steps where files are written to disk and then combined using only a single core. It means that we also automatically have one {\tt les houches event} file for each core, which can then be fed to \hackanalysis for the showering and analysis stage. Moroever, the running of the gridpacks can also be performed on a ramdisk (typically using {\tt /dev/shm} on linux systems; this is the default location for the \BSMArt temporary directories when not running in debug mode), provided that not too many events are simulated in one go: this means we can almost completely eliminate disk reading/writing operations, which is often a major drag on the analysis chain.

To take advantage of this possibility, we should use of batches of events, that will be run through \hackanalysis and then the outputs combined. This enables the possibility of checking for convergence and stopping the running of batches when some criterion is met. There are two possibilities:
\begin{enumerate}
\item Halting the simulation when the uncertainty on the efficiency of each region is within a specified margin (say $10\%$), for regions above a specified minimum efficiency. This is appropriate when generating cutflows and developing the analysis. 
\item Comparing to the experimental data. We halt the simulation when all regions have converged (according to the following criteria):
  \begin{enumerate}
  \item The uncertainty for the region is below the experimental sensitivity.
  \item The upper bound estimate for the efficiency of the region is below the experimental sensitivity (i.e. too few events would be generated).
  \item The lower bound estimate for the efficiency of the region is above the level required to exclude the parameter point by itelf (i.e. we easily see that too many events are generated). 
    \end{enumerate}
  \end{enumerate}

  The implementation of the first criterion is activated by setting {\tt Check Absolute Convergence} to {\tt true} in the json file. The relevant settings, given here for the example used to generate the cutflows in tables \ref{TAB:LM}, \ref{TAB:MM}, \ref{TAB:HM}, are:
\begin{lstlisting}[language=HackAnalysis,title=absolute convergence checks json settings]
{
   "Codes":{
        "MadGraphHackAnalysis":{
            "InputFile": "chiwh.slha",
            "HackAnalysis": "/path/to/HackAnalysis_v2.1/analysePYTHIA_LHE.exe",
            "MadGraph Path": "/path/to/MG5/bin/mg5_aMC",
            "GridPack": "True", 
            "python2": "False",
            "Batch Size": 40000,
            "Check Absolute Convergence": "True",
	    "Convergence Margin": 0.1,
	    "Efficiency Threshold": 1e-3,
            "Cores": 4,
            "Proc Card": "chiWH.proc",
             "Process": "ppCN",
             "YAML file": "chiwh.yaml",
             "Config file": "chiwh.cfg",
             "Store Events": "False",
             "Efficiency Filename": "chiwh.eff",
             "Cutflow Filename": "chiwh_cf.eff",
            "Histogram Filename": "chiwh.yoda",
             "Run": "True",
             "Observables": { "xs": {"SLHA":["MADGRAPH",[1]]}}
           }
         },
         ...
}
\end{lstlisting}
The most relevant settings are "Convergence Margin" of $10\%$ and "Efficiency Threshold" giving the minimum efficiency to consider necessary for convergence. In the simplest scans (as above), the \emph{only} code invoked is {\tt MadGraphHackAnalysis}, and then the {\tt InputFile} (here {\tt chiwh.slha}) for all scans except {\tt read\_dir} must already exist in the launch directory as a template from which to generate the parameter cards.

Note that the maximum number of events to generate is not specified here -- it is read from the setting {\tt nevents} in the \hackanalysis yaml file:
\begin{lstlisting}[language=HackAnalysis,title=chiwh.yaml]
---
analyses:
 - ATLAS_SUSY_2019_08
settings:
  nevents: 1200000
  debug: true
  cores: 4
  Efficiency Filename: chiwh.eff
  Cutflow Filename: chiwh_cf.eff
  Histogram Filename: chiwh.yoda
  Include Pileup: false
  LHE file: Split/split
  Config file: chiwh.cfg
\end{lstlisting}
In the example tables provided below, although a maximum of $1.2M$ events was specified, no point required as many events as that; only about $200k$ were needed each time, and this is the virtue of the convergence technique.
The LHE file setting is read from this file and used to generate the relevant files for each core; the default is {\tt Split/split} which means that it will create a subdirectory {\tt Split} and populate it with files {\tt Split/split\_0.lhe.gz,Split/split\_1.lhe.gz, ...} etc.  

Note that the {\tt Proc Card} setting in the json file is used to generate the \madgraph model directory during initialisation; it is a file containing a series of commands, in this case:
\begin{lstlisting}[language=SLHA,title=chiWH.proc]
import model MSSM_SLHA2 -modelname
define HN = n2 
define CHAP = x1+ 
define CHAM = x1-
define exc = go h01 h02 a0 h+ h- n1 n2 n3 n4 x1+ x2+ x1- x2- sv1 sv2 sv3 sl1- sl2- sl3- sl4- sl5- sl6- sl1+ sl2+ sl3+ sl4+ sl5+ su1 su2 su3 su4 su5 su6 sd1 sd2 sd3 sd4 sd5 sd6 su1~ su2~ su3~ su4~ su5~ su6~ sd1~ sd2~ sd3~ sd4~ sd5~ sd6~
generate p p > HN CHAP /exc
add process p p > HN CHAM /exc
add process p p > HN CHAP j /exc
add process p p > HN CHAM j /exc 
add process p p > HN CHAP j j /exc 
add process p p > HN CHAM j j /exc
output MSSMChiWH
\end{lstlisting}
\BSMArt will automatically read the name of the output directory from this file.  
  
For the second method, we check each region in turn, provided that the statistical information about the analysis is placed in the {\tt Statistics} subdirectory of \hackanalysis (see section \ref{SEC:STATS}): for a new analysis, in order to take advantage of this feature, the relevant json files must be created. If we find that criterion (c) is satisfied for \emph{any} region, then we stop checking and declare all regions converged -- there is no need to simulate more events if we are sure that the parameter point is excluded. In general, this procedure should save a large amount of time and computing power: for the typical case of trying to reproduce an experimental exclusion curve, the traditional approach of demanding a fixed number of events per parameter point is vastly inefficient for the regions away from the exclusion curve where the cross-section/efficiencies are too low or too high, and it should not be necessary to simulate a large number of events in those regions. 

In order to invoke these routines, the settings are idential to the above, except we select {\tt Check Convergence} to be true instead of {\tt Check Absolute Convergence}, and the settings  "Convergence Margin" and "Efficiency Threshold" are ignored:
\begin{lstlisting}[language=HackAnalysis,title=exclusion convergence checks json settings]
{
    "Codes":{
        "MadGraphHackAnalysis":{
            "InputFile": "chiwh.slha",
            "HackAnalysis": "/path/to/HackAnalysis_v2.1/analysePYTHIA_LHE.exe",
            "MadGraph Path": "/path/to/MG5/bin/mg5_aMC",
            "GridPack": "True", 
            "Batch Size": 40000,
            "Check Convergence": "True",
...
\end{lstlisting}

In the future the convergence checking/running of batches of events will be extended to the other modes of running \hackanalysis -- and also to running \madanalysis through \BSMArt.

\section{New physics and mathematics routines: RestFrames, Eigen, MT2, Nelder-Mead minimisation}
\label{SEC:RESTFRAMES}

In order to recast the analysis ATLAS-SUSY-2018-16 \cite{ATLAS:2019lng} it was necessary to have an implementation of the {\tt restframes} package \cite{Jackson:2017gcy}\footnote{\url{http://restframes.com/}}. As outlined in \cite{Agin:2024yfs}, this required replacements for many {\tt root} functions for vector/matrix manipulation and function minimisation. As a result, additions to the {\tt heputils} $\rightarrow$ {\tt HEP} namespace functions were made, including three-vectors, rotations around a fixed axis, and including the {\tt Eigen3} package\footnote{\url{https://eigen.tuxfamily.org/}; it is not necessary to download the code separately, {\tt Eigen3} is included with \hackanalysis (for now).}. It would be tedious and lengthy to describe in detail all of these functions, but the user can consult the code in the {\tt include/ExHEPUtils}, {\tt include/RestFrames}, {\tt include/Eigen} and {\tt include/MATHS} subdirectories of the installation, or the doxygen documentation provided, which will become an ever more comprehensive reference. Alternatively the reader can consult the source code for the relevant analysis.

Invoking {\tt restframes} is highly straighforward via the helper functions modified from {\tt SimpleAnalysis} \cite{ATLAS:2022yru}, which makes the resulting code as close to the ATLAS pseudocode as possible. Similarly, the MT2 variable \cite{Lester:1999tx,Barr:2003rg,Lester:2014yga} is provided via the bundled header file {\tt include/lester\_mt2\_bisect.h} automatically included in the {\tt BaseAnalysis.h} header (so no need to specify it) which provides the function {\tt asymm\_mt2\_lester\_bisect::get\_mT2}.

\section{Linking ONNX}
\label{SEC:ONNX}

An increasingly common feature of new LHC analyses is the use of machine learning to separate signal from background. In line with best-practice recommendations \cite{Araz:2023mda} the trained networks can be made public by the collaborations, and perhaps the best or most widely useable of the formats for this is ONNX\footnote{\url{https://onnx.ai/}} \cite{Araz:2023mda}. ONNX provides easily-installable libraries, so that the training can be performed in any language or platform and then imported in any language or platform. Relevant for \hackanalysis then are the \cpp libraries: once the user has installed these on their system (an installer is not provided for this, but the installation is straightforward from the ONNX website), \hackanalysis can be linked to ONNX easily by just adding the path to the libraries in {\tt Paths.inc} -- as can be seen in section \ref{SEC:Installation}.

To actually use ONNX in a recast or new analysis there are helper functions which automate the whole thing. These are contained in the header file {\tt include/handle\_onnx.h}. A class {\tt HEPONNX} is defined which contains the method {\tt addonnx} which takes as a single argument a cstring which is the path to the relevant ONNX file (an exported network). These are supposed to be contained in the {\tt DATA} subdirectory of the \hackanalysis installation, so once the class is declared in the analysis header files, e.g.:
\begin{lstlisting}[language=C++,title=example ONNX analysis header file]
#ifdef ONNXPATH
#ifndef ATLAS_SUSY_2019_04_h
#define ATLAS_SUSY_2019_04_h
#include "include/BaseAnalysis.h"
#include "include/HEPData.h"
#include "include/handle_onnx.h"

using namespace std;

class ATLAS_SUSY_2019_04 : public BaseAnalysis {
    public:

       ~ATLAS_SUSY_2019_04();
       void init();
       void Execute(std::mt19937 &engine);
       void Finalise() { return; }
       ATLAS_SUSY_2019_04() { this->setup(); this->analysisname="ATLAS_SUSY_2019_04";  };
        HEPONNX onnx4jets,onnx5jets,onnx6jets,onnx7jets,onnx8jets;
 

};
#endif
#endif
\end{lstlisting}
then the networks can be initialised in the {\\ init()} function of the analysis:
\begin{lstlisting}[language=C++]
void ATLAS_SUSY_2019_04::init() {

this->DetectorFunction="DEFAULT";

onnx4jets.addonnx(string("ATLAS_SUSY_2019_04/SUSY-2019-04_4jets.onnx"));
onnx5jets.addonnx(string("ATLAS_SUSY_2019_04/SUSY-2019-04_5jets.onnx"));
onnx6jets.addonnx(string("ATLAS_SUSY_2019_04/SUSY-2019-04_6jets.onnx"));
onnx7jets.addonnx(string("ATLAS_SUSY_2019_04/SUSY-2019-04_7jets.onnx"));
onnx8jets.addonnx(string("ATLAS_SUSY_2019_04/SUSY-2019-04_8jets.onnx"));
\end{lstlisting}
and then invoked by:
\begin{lstlisting}[language=C++]
  void ATLAS_SUSY_2019_04::Execute(std::mt19937 &engine) {

    ...
  
  vector<float> nninputs;

  ...
  
 float nn_value=onnx4jets.get_result(nn_inputs);
\end{lstlisting}
The inputs/outputs to the ONNX routines are floats rather than double-precision numbers.

All of the technical details of the input/output are then abstracted and it is entirely straightforward to add these to analyses. I give no examples for the output here, as it is intended to present details of recasting the analysis ATLAS-SUSY-2019-04 elsewhere along with a madanalysis implementation.

\section{New included analyses}
\label{SEC:ANALYSES}

With version 2 the list of included analyses is:
\begin{itemize}
\item {\tt HSCP\_ATLAS}  \cite{ATLAS:2019gqq,Goodsell:2021iwc}.
\item {\tt DT\_CMS} \cite{CMS:2018rea,CMS:2020atg,Goodsell:2021iwc}.
\item {\tt ATLAS\_SUSY\_2017\_04\_2body}, {\tt ATLAS\_SUSY\_2017\_04\_3body} \cite{ATLAS:2019fwx,Araz:2021akd}.
\item {\tt ATLAS\_SUSY\_2019\_08} \cite{ATLAS:2020pgy,Goodsell:2020ddr,Fuks:2021zbm}.
\item {\tt ATLAS\_EXOT\_2018\_06} \cite{ATLAS:2021kxv}, based on the recast by D.~Agin on the \madanalysis PAD.
\item {\tt CMS\_EXO\_20\_004} \cite{CMS:2021far}, based on the recast on the \madanalysis PAD.
\item {\tt ATLAS\_SUSY\_2018\_16} \cite{ATLAS:2019lng,Agin:2024yfs}.
\item {\tt ATLAS\_SUSY\_2019\_09\_onshell}, {\tt ATLAS\_SUSY\_2019\_09\_offshell} \cite{ATLAS:2021moa,Agin:2024yfs}.
\end{itemize}
This is (still) relatively modest: the goal of the code, however, is not to be a repository but to be used for development, and whenever a fast and accurate recast is needed for a given task. Yet these comprise a relatively comprehensive set of analyses for electroweakino searches; the most powerful searches for most regions of the parameter space are present, with the exception of when the mass splittings become very large (hopefully this will be remedied in future).

The long-lived particle analyses {\tt HSCP\_ATLAS}, {\tt DT\_CMS} were included in the first version of \hackanalysis and their validation was described in \cite{Goodsell:2021iwc}. The displaced lepton search {\tt ATLAS\_SUSY\_2017\_04} was implemented in parallel with the implementation in \madanalysis described in \cite{Araz:2021akd} and first appeared in \hackanalysis version 1.2. The others represent new additions with version 2 and are discussed more below.

\subsection{ATLAS-SUSY-2019-08}

The recast of the electroweakino search {\tt ATLAS\_SUSY\_2019\_08} was originally written in a prototype version of \hackanalysis as an aid to the development of the analysis \cite{Goodsell:2020ddr,Fuks:2021zbm} in \madanalysis; it has now been updated and included with the release of \hackanalysis. Since the implementation in \madanalysis was required \delphes, this latest version could in principle lead to an SFS code. 

The search targets the process $p p \rightarrow \tilde{\chi}_2^0 \tilde{\chi}_1^\pm$ where $\tilde{\chi}_2^0$ decays to $\tilde{\chi}_1^0$ plus an on-shell Higgs boson, which in turn decays to b-jets; while the $\tilde{\chi}_1^\pm$ should decay to the lightest neutralino plus an on-shell $W$ boson, which then decays leptonically. In the interpreted scenario  $\tilde{\chi}_2^0$ and $\tilde{\chi}_1^\pm$ are degenerate. The simulated events correspond to the above process with up to two additional partons in the final state and with the decays forced and handled through \pythia. 
To validate the code I show the cutflows in tables \ref{TAB:LM}, \ref{TAB:MM}, \ref{TAB:HM} for the corresponding points $m_{\tilde{\chi}_2^0}/m_{\tilde{\chi}_1^\pm}, m_{\tilde{\chi}_1^0} = (300 ,75 ), (300 ,75 ) $ and $(750, 100) $ GeV. The agreement between the ATLAS cutflows and \hackanalysis is excellent; in addition, these cutflows show the systematic uncertainties computed automatically by \hackanalysis and were produced using the \BSMArt machinery described in section \ref{SEC:BSMART}.

\subsection{ATLAS-SUSY-2018-16 and ATLAS-SUSY-2019-09}

These are soft lepton searches where an excess is seen in the data; they are nevertheless the most powerful searches for compressed electroweak-charged particles. The recasting and validation by means of reproducing the ATLAS exclusion curves is described in \cite{Agin:2024yfs}. To supplement the material there, in tables \ref{TAB:1L1T}, \ref{TAB:SREhigh}, \ref{TAB:SRElow} and \ref{TAB:SREmed} I show a cutflow comparison for the search ATLAS-SUSY-2018-16; ATLAS provided only a cutflow for higgsinos with masses $(m_{\tilde{\chi}_2^0},m_{\tilde{\chi}_1^\pm},m_{\tilde{\chi}_1^0}) = (155, 152.5,150)$ GeV for the process $p p \rightarrow \tilde{\chi}_2^0 \tilde{\chi}_1^0/\tilde{\chi}_2^0 \tilde{\chi}_1^\pm$ plus up to two jets in the final state. I generate these as two sets of processes processes  and merge the cutflows (which can be performed using the scripts provided with \hackanalysis) weighted appropriately by their cross-sections. The decay $\tilde{\chi}_2^0 \rightarrow \ell^+ \ell^- \tilde{\chi}_1^0$ is forced in the the hard event in \madgraph and therefore the full matrix element is computed, which, as described in \cite{Agin:2024yfs}, is necessary to correctly model the lepton pair invariant mass distribution. The cutflows show excellent agreement between \hackanalysis and the experimental data.

\subsection{ATLAS-EXOT-2018-06 and CMS-EXO-20-004}
\label{sec:monojets}

These monojet searches were already implemented in the \madanalysis PAD (although the ATLAS search was implemented in parallel in \hackanalysis during its development as a cross-check) and were ported to \hackanalysis for the purposes of comparing the excesses in the soft-lepton search with those in the monojet searches discussed in \cite{Agin:2023yoq,Agin:2024yfs}.

\begin{figure}[!h]\centering
  \includegraphics[width=0.7\textwidth]{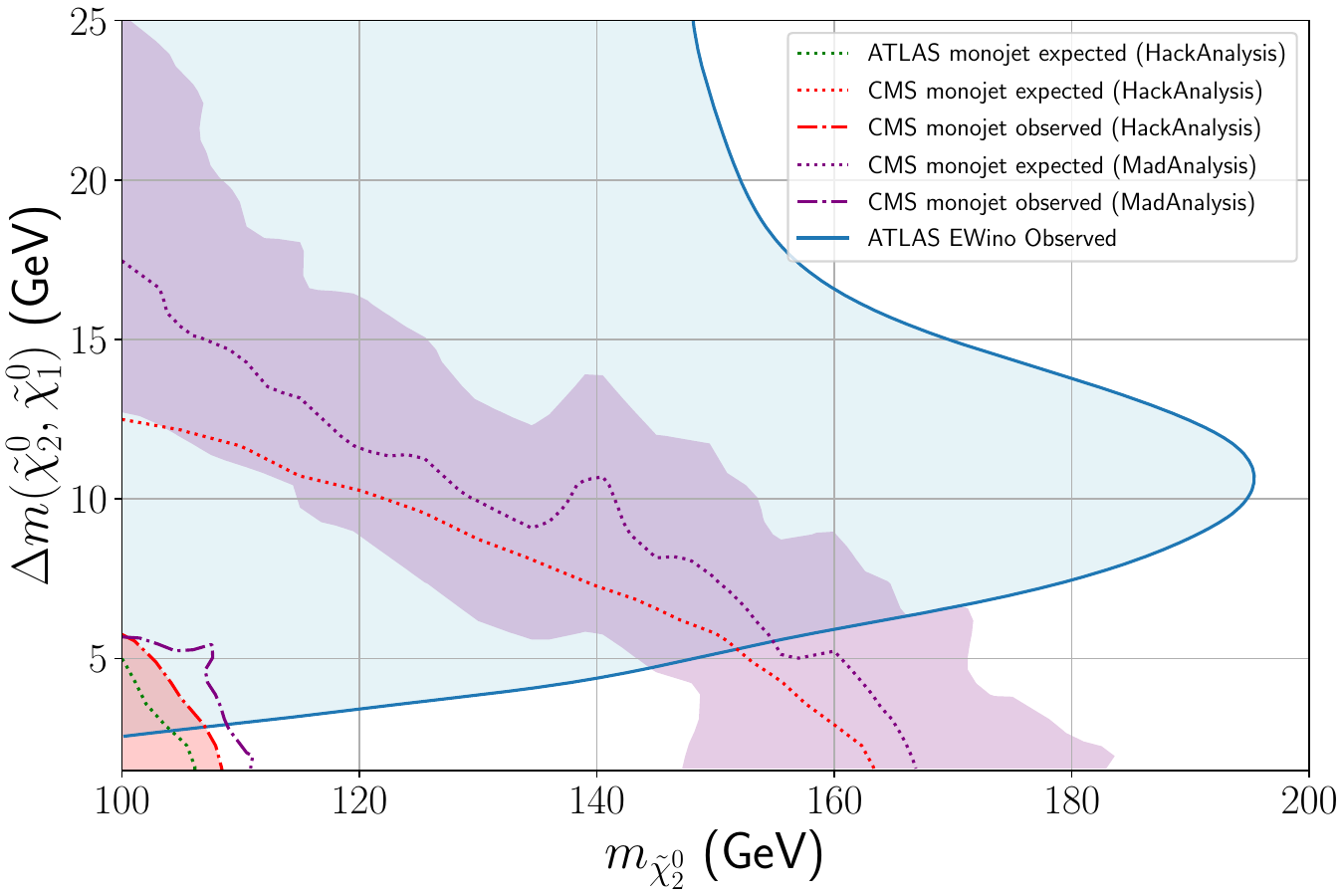}
 \caption{\label{FIG:Monojet} Comparison of monojet limits for higgsinos between \hackanalysis and the results of  \madanalysis from \cite{Agin:2023yoq}. The purple band around the CMS monojet expected limit in \madanalysis shows the region obtained by varying the cross-section by  $\pm 10\%$; this shows that the modest differences in the curves are within uncertainties. }
\end{figure}

As an example of the added power available using the gridpack machinery, which ennables a large number of events to be generated, figure \ref{FIG:Monojet} shows a comparison between the expected and observed limits from this analysis upon pure higgsinos with masses split such that $m_{\tilde{\chi}_1^\pm} = \frac{1}{2} ( |m_{\tilde{\chi}_1^2}| + |m_{\tilde{\chi}_1^0}|)$ and all other supersymmetric particles are decoupled. Events are simulated for production of all of pairs of electroweakinos with up to two additional partons and the decays are performed in \pythia. Running \hackanalysis with gridpacks on a cluster where from each node 8 cores are requested, and  six batches of 3.2M events per parameter point leads to 19.2M events per point, where the batches are run in parallel and take roughly four hours each to be generated and analysed. Compare this to running \madanalysis for the results in \cite{Agin:2023yoq} where up to 8M events were generated per parameter point on essentially the same architecture, but in batches of 2M events, each of which took around 7 hours to generate the \hepmc files (split into 40 batches) and a further nine hours to perform the analysis (running through both ATLAS-EXOT-2018-06 and CMS-EXO-20-004) making 16 hours total for 2M events; the workflow for \hackanalysis is therefore at least six times faster, and potentially more because with the nodes being occupied for much smaller amounts of time (so can fit more easily into a busy batch schedule) and much smaller hard-drive usage, so more can be run in parallel. Hence overall it was possible to simulate more points in a reasonable time, and obtain a denser grid from which to construct a smooth contour. Moreover, this example does not use the batch mode in \BSMArt, instead writing the {\tt les houches events} to disk in one step; using batches and ramdisk would lead to a further performance enhancement. 

\begin{figure}[!h]\centering
  \includegraphics[width=0.7\textwidth]{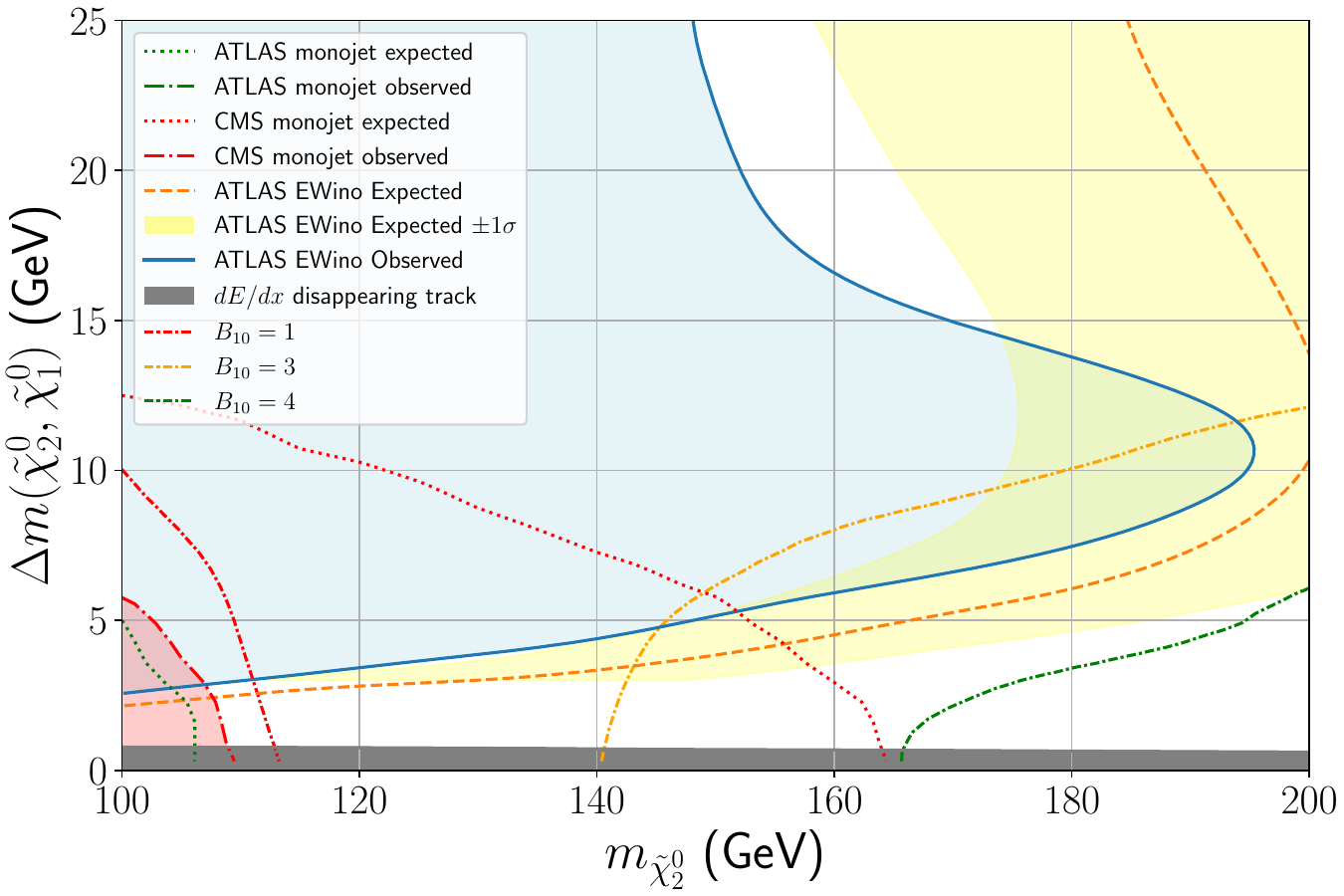}
 \caption{\label{FIG:MonojetBayes} Monojet expected and observed limits computed using \hackanalysis, with contours of constant Bayes Factor $B_{10}$ (see text for details)}
\end{figure}

\section{Statistics routines}
\label{SEC:STATS}

Once an analysis has been checked against cutflows, in order to use it the outputs must be checked against experimental data to test either for exclusion -- or for significance, in the case of excesses. Each recasting tool has its own routines for single-bin (or even mulit-bin) statistical checks, but otherwise for more sophisticated statistical models it is necessary to use {\tt pyhf} \cite{Heinrich:2021gyp} and/or {\tt spey} \cite{Araz:2023bwx}. It is then tedious to write code to process the efficiency information and handle the running of these {\tt python} packages; so with \hackanalysis routines are provided to do just that. In the {\tt Statistics} directory of the installation are numerous scripts for handling routine tasks, such as combining cutflows, writing cutflows in \LaTeX\  format, and the {\tt ha\_convergence.py} script for checking convergence. As a subdirectory there is a {\tt python} package {\tt HAStats} for handling all the statstical treatment, and a script for invoking it, when supplied with a \hackanalysis {\tt SLHA}-style efficiency file or json file:
\begin{lstlisting}[]
python3 HAStats.py <HA efficiency file OR json file> [Optional: --xsection=] [Optional: --infopath=] [Optional: --exclusion] 
\end{lstlisting}
This will load the relevant statistics handler ({\tt pyhf}, {\tt spey} or {\tt Native}) and run it, computing exclusion limits and significances, unless the option {\tt exclusion} is selected, in which case only the limits are computed.

The packages can of course be imported into a {\tt python} script in order to run customised checks. There is also a script {\tt BSMArtHA.py} which will run exclusion checks on all of the outputs of a \BSMArt scan when given a \BSMArt json file as input. It is intended that (near) future \BSMArt releases will include the option to run the statistics module of \hackanalysis to compute exclusion bounds and significances for both \hackanalysis and \madanalysis during the running of the code: for the time being, the full statistics checks (as opposed to the convergence checks) are only run after a scan is complete. 

As an example of the possibilities for statistical interpretation, in figure \ref{FIG:MonojetBayes} I show again the expected and observed limits from \hackanalysis for the ATLAS and CMS monojet searches presented in section \ref{sec:monojets} and figure \ref{FIG:Monojet}, but this time with the contours of constant \emph{Bayes Factor} $B_{10}$ for the CMS monojet search. As mentioned above, this search contains excesses, and yet the standard way of presenting information about excesses is to show the difference between expected and observed limits, and to give $p$-values for benchmark points. However, neither of these quantities show the region where \emph{a specific model} agrees best with data; the $p$-value is computed comparing the Standard Model with the best-fit value of the signal. On the other hand, the Bayes Factor is the ratio of evidences for the model and the Standard Model; in the figure, instead of the full quantity, $B_{10}$ is computed as the ratio of likelihoods between signal plus background and background-only models, as advocated in section 3.2 of \cite{Fowlie:2024dgj}. The likelihoods are computed using {\tt spey} and the simiplified-likelihood information provided by CMS. The contour of $B_{10} =1 $ should roughly correspond to the exclusion contour, while the regions of increasing $B_{10}$ indicate better agreement of the model with the data: it can be seen therefore that, surprisingly, the CMS monojet search is better fit by larger $m_{\tilde{\chi}_2^0}$ and unsurprisingly smaller $\Delta m$, yet the region that would correspond to the soft lepton excess still has $B_{10} \in [1,3]$.

\section{Writing your own analysis}
\label{SEC:YOURANALYSIS}

The first step in creating a new analysis is just to run the provided {\tt python} script to generate the analysis \cpp and header files, and update the {\tt analysislist.h} header file which tells the code which analyses are included:
\begin{lstlisting}[]
./newAnalysis [Optional: --ONNX] NewAnalysisName
\end{lstlisting}
This creates the files {\tt NewAnalysisName.cc} and {\tt NewAnalysisName.h} in the directory {\tt UserAnalyses}, with the basic content:
\begin{lstlisting}[language=HackAnalysis,title=MyNewAnalysis.cc]
#include "MyNewAnalysis.h"

namespace HEP
{
    namespace MyNewAnalysis 
    {
/* Place here your user-defined functions */

    }
}

using namespace HEP::MyNewAnalysis;

void MyNewAnalysis::init() {

cout << "---------------------------------------------------" << endl;
cout << "-- Analysis MyNewAnalysis --" << endl;
cout << "---------------------------------------------------" << endl;
}

MyNewAnalysis::~MyNewAnalysis() {

};

void MyNewAnalysis::Execute(std::mt19937 &engine) {

}
\end{lstlisting}
and
\begin{lstlisting}[language=HackAnalysis,title=MyNewAnalysis.h]
#ifndef MyNewAnalysis_h
#define MyNewAnalysis_h
#include "include/BaseAnalysis.h"
#include "include/HEPData.h"

using namespace std;

class MyNewAnalysis : public BaseAnalysis {
    public:

       ~MyNewAnalysis();
       void init();

       void Execute(std::mt19937 &engine);

       void Finalise() { return; }

       MyNewAnalysis() { this->setup(); this->analysisname="MyNewAnalysis"; };

};

#endif
\end{lstlisting}

If the {\tt --ONNX} flag is added, the appropriate header files are linked and definitions added so that {\tt ONNX} networks can be included.

Typically it is only necessary to add content to the {\tt init()} and {\tt Execute} functions in the \cpp file. For the {\tt init()} function, that is where the user should update the fill function (as seen in examples above) if needed, and define the cutflows. The syntax is similar to that of \madanalysis: signal regions are defined by {\tt AddSignalRegion(<regionname>)} and cuts in one or multiple regions are defined by {\tt AddCut(<cutname>,<regionname>)} or {\tt AddCut(<cutname>,<vector of cuts>)}. E.g. some lines from the ATLAS-SUSY-2019-08 analysis:
\begin{lstlisting}[language=HackAnalysis]
void ATLAS_SUSY_2019_08::init() {

AddRegionSelection("LMdisc"); 
AddRegionSelection("MMdisc"); 
AddRegionSelection("HMdisc"); 
 
...

const std::vector<std::string> alldiscSRs={"LMdisc","MMdisc","HMdisc"}; 

 
AddCut("Njets25 >=2",allSRs); 
AddCut("1 signal lepton",allSRs); 

...
AddCut("mT > 100","LMdisc"); 
AddCut("mT > 160","MMdisc"); 
 
AddCut("minCT",alldiscSRs); 
...
this->DetectorFunction="ATLAS";
}
\end{lstlisting}
In the {\tt Execute} function, booking of cuts is performed by the void function {\tt ApplyCut(bool condition,cutname)} which will automatically add the weights remaining to the relevant cut in every cutflow that contains it, e.g.
\begin{lstlisting}[language=C++]
bool IsLeptons = (nleptons == 1); 
bool IsNJets = (NJets < 4);	   
...	 
ApplyCut(IsNJets2,"Njets25 >=2"); 
ApplyCut(IsLeptons,"1 signal lepton"); 
\end{lstlisting}

The functionality is deliberately very similar to \madanalysis but the running is similar to \GAMBIT, in order to facilitate easy inclusion in the more popular codes after an analysis has been developed (and indeed there are perl scripts for converting as close to \madanalysis syntax as possible).

Note in the above that the use of namespaces for each analysis allows the user to define their own functions without fear of name clashes. However, if it is desired to define functions shared accross analysis, it is also possible to put these  in the {\tt UserAnalyses} directory: if the corresponding header files have extension {\tt .hpp} they will be picked up for building but not treated as analyses. 

In order to delete an analysis, all the user has to do is delete the {\tt .h} and {\tt .cc} files, and run the script:
\begin{lstlisting}[]
./regenerate.py
\end{lstlisting}
which will clean up the {\tt analysislist.h} header file.

\subsection{Booking histograms and adding efficiency functions}

Adding histograms to analyses is very simple: it is only necessary to declare the histogram in the {\tt init()} function using {\tt AddYodaHisto1D(<name>,<edges>)}, e.g.
\begin{lstlisting}[language=C++]
  void DT_CMS::init() {
    ...
 vector<double> METedges={10,30,50,70,90,110,130,150,200,250,300,350,400,1000.0};
    
AddYodaHisto1D("MET",METedges);
AddYodaHisto1D("pTJ1",100,10.0,2000.0);
\end{lstlisting}
and then they can be filled inside the event via {\tt FillYodaHisto1D(<name>,<value>)}:
\begin{lstlisting}[language=C++]
FillYodaHisto1D("MET",MET);
if( Event->jets().size() > 0)
  {
    FillYodaHisto1D("pTJ1", Event->jets()[0]->pT());

  }
\end{lstlisting}
Two-dimensional histograms can be added via {\tt AddYodaHisto2D}; these functions are defined in {\tt include/BaseAnalysis.h}. The histograms generated in each core during running are added and written to file at the end. In order to analyse and plot the histograms, {\tt yoda} contains many useful helper tools such as {\tt yoda-plot}; these are very user-friendly and examples of histograms produced from these tools are to be found in \cite{Agin:2024yfs}, otherwise the reader is referred to the {\tt yoda} documentation. As an aside, if the user has used the install scripts provided to install {\tt yoda}, then the relevant {\tt python} paths can be configured using {\tt SETUP\_COLLIDERS.sh} or {\tt configure\_colliders.csh} scripts. 

Efficiency functions can be implemented via {\tt heputils} binning functions. These are wrapped into classes {\tt Efficiency1D}, {\tt Efficiency2D} and {\tt YodaEfficiencies} in the file {\tt include/HEPData.h}, which is used for reading efficiencies from {\tt .csv} and {\tt .yoda} files provided on \href{http://hepdata.net}{HEPData}. To use these, the user can declare them as class members in the header file, e.g.
\begin{lstlisting}[language=C++,title=HSCP\_ATLAS.h]
...
class HSCP_ATLAS : public BaseAnalysis {
    public:
    ~HSCP_ATLAS();
    void init();
    void Execute(std::mt19937 &engine);
    void Finalise() { return; }   
    HSCP_ATLAS() { this->setup(); this->analysisname="HSCP_ATLAS";  };

   Efficiency1D EtmissTurnOn;   
   Efficiency2D SingleMuTurnOn;
...
\end{lstlisting}
They can then be initialised and used:
\begin{lstlisting}[language=C++,title=HSCP\_ATLAS.cc]
  ...
void HSCP_ATLAS::init() {
  this->DetectorFunction="ATLAS";
...
EtmissTurnOn.readcsv("HSCP-ATLAS/Table22.csv");   
SingleMuTurnOn.readcsv("HSCP-ATLAS/Table23.csv");
...
}
...
void HSCP_ATLAS::Execute(std::mt19937 &engine) {
  ...
double eff_Met = EtmissTurnOn.get_at(Etmiss);
double effTrig   = SingleMuTurnOn.get_at(abseta, beta);
...
\end{lstlisting}
The function {\tt readcsv} will look for tables in the {\tt DATA} directory of the \hackanalysis installation.

\subsection{Adding statistical information}

Finally, in order to take advantage of the statistics routines for a new analysis, it is necessary to create a json info file with the analysis information and place it in the {\tt Statistics/HAStats/Analyses} file. This should contain as a minimum the analysis name, luminosity, analysis type -- which can be {\tt pyhf, MA5spey, Native} -- and data about the expected and observed events in {\tt RegionData} (even for {\tt pyhf} and {\tt MA5spey} analyses); for {\tt pyhf} and {\tt MA5spey} analyses a {\tt Background File} is needed. By {\tt Native} it is understood that we mean that the statistics should be handled by the native code included with \hackanalysis, which means single bins or bins with no correlations between them; {\tt MA5spey} means a simplified likelihood with correlations as specified in a \madanalysis ``{\tt .info}'' file. For So as an example we have:
\begin{lstlisting}[language=HackAnalysis,title=ATLAS\_SUSY\_2018\_16\_info.json]
{
    "Analysis": "ATLAS_SUSY_2018_16",
    "Lumi": 139.0,
    "Type": "pyhf",
    "Background File": "ATLAS_SUSY_2018_16_bkg.json",
    "Region Dictionary": {
        "SR_eMLLa_Onelep1track_cuts": "SR_E_lT_eMLLa",
        "SR_eMLLb_Onelep1track_cuts": "SR_E_lT_eMLLb",
   ...
    },
    "RegionData": {
        "SR_E_lT_eMLLa" : {
            "nb":0.5,
            "deltanb":0.5,
            "observed": 0
        },
        "SR_E_lT_eMLLb": {
            "nb":6.0,
            "deltanb":1.9,
            "observed": 8
        },
 ...
}

\end{lstlisting}
For MyNewAnalysis it should be named {\tt MYNEWANALYSIS\_info.json} (with the analysis name all in upper case, and {\tt \_info.json} appended). The {\tt RegionData} lists the number of expected background events ({\tt nb}), uncertainty on the background ({\tt deltanb}) and number of events actually observed ({\tt observed}). For {\tt Native} analyses this information is used to compute the exclusion limits by checking the apriori and observed $CL_s$ values for each bin, and taking the observed value from the bin with the best apriori value. In addition, this information is used for every type of analysis  in convergence checks, to set the experimental sensitivity and to see whether a single bin excludes the point. 
For analyses with a {\tt pyhf} background file, this should be placed in the {\tt Statistics/HAStats/Data} directory, and it is necessary to give a dictonary to map the cuts as they appear in the analysis code to the names in the background file. To aid this, a script {\tt process\_pyhf.py} is provided: running this after specifying the background file and the name of the analysis will generate the necessary json information file equivalent to the one above that can be edited and placed in the {\tt Statistics/HAStats/Analyses} directory. For {\tt MA5spey} analyses, meaning that we use a simplified likelihood, the background file is a \madanalysis {\tt .info} file; processing these can similarly be done via the script {\tt process\_madanalysis\_info.py}.

\section{Outlook}
\label{SEC:OUTLOOK}

It is hoped that \hackanalysis can aid in the development of new features for recasting and proposing new analyses due to its flexibility and speed. Ideally users will add their own features and then propose them for additions either to \hackanalysis itself or to one of the major recasting tools. While it would also be desirable to collect contributed analyses for \hackanalysis, it is not the intention to create another framework competing with the others, but instead to aid interoperability and work towards the goal that as many LHC analyses are recast as possible.

As mentioned above, further integration of \hackanalysis with \BSMArt for scanning parameter spaces and checking exclusion or compatibility of models with data is intended. At the same time, the techniques for running batches of points and checking for convergence should be extended to running \madanalysis, and it is hoped that by collaboration with the authors that more efficient and economical ways of taking advantage of its large library of analyses (and contributing to them) will be possible.



\section*{Acknowledgements}

MDG acknowledges support from the grant \mbox{``DMwithLLPatLHC''} of the Agence Nationale de la Recherche
(ANR) (ANR-21-CE31-0013). I thank Lakshmi Priya for collaboration during development of version 1. I thank Diyar Agin, Benjamin Fuks and Taylor Murphy for collaboration on projects using \HackAnalysis. I thank Benjamin Fuks and Taylor Murphy for comments on the draft. I thank in addition Sabine Kraml for helpful discussions.

\appendix

\section{Cutflows}

\begin{table}\small\centering
\begin{tabular}{|c||c|c||} \hline\hline
Cut & ATLAS & HackAnalysis \\ \hline\hline
All weighted events & $ 1.0 $ & $ 1.0^{+0.00}_{-0.00} $ (stat) ${}^{+ 9.3\%}_{-7.4\%}$ (syst)\\
$N_{\mathrm{jets,25}} \geq 2$ & $ 8.1\times 10^{-1} $ & $ 8.5^{+0.01}_{-0.01}\times 10^{-1} $ (stat) ${}^{+ 1.1\%}_{-1.0\%}$ (syst)\\
1 signal lepton & $ 7.0\times 10^{-1} $ & $ 7.0^{+0.01}_{-0.01}\times 10^{-1} $ (stat) ${}^{+ 1.1\%}_{-1.1\%}$ (syst)\\
Second baseline lepton veto & $ 6.9\times 10^{-1} $ & $ 7.0^{+0.01}_{-0.01}\times 10^{-1} $ (stat) ${}^{+ 1.1\%}_{-1.1\%}$ (syst)\\
$m_{\mathrm{T}} > $ 50 GeV & $ 5.6\times 10^{-1} $ & $ 5.6^{+0.02}_{-0.02}\times 10^{-1} $ (stat) ${}^{+ 1.1\%}_{-1.1\%}$ (syst)\\
$E_\mathrm{T}^\mathrm{miss} > $ 180 GeV & $ 1.6\times 10^{-1} $ & $ 1.6^{+0.01}_{-0.01}\times 10^{-1} $ (stat) ${}^{+ 4.1\%}_{-4.0\%}$ (syst)\\
$N_{\mathrm{jets}} \leq 3$ & $ 1.4\times 10^{-1} $ & $ 1.2^{+0.01}_{-0.01}\times 10^{-1} $ (stat) ${}^{+ 1.0\%}_{-1.4\%}$ (syst)\\
$N_{\mathrm{b-jets}}=2$ & $ 5.7\times 10^{-2} $ & $ 5.3^{+0.07}_{-0.07}\times 10^{-2} $ (stat) ${}^{+ 2.1\%}_{-2.3\%}$ (syst)\\
$m_{\mathrm{bb}}$ $>$ 50 GeV & $ 5.7\times 10^{-2} $ & $ 5.2^{+0.07}_{-0.07}\times 10^{-2} $ (stat) ${}^{+ 2.1\%}_{-2.3\%}$ (syst)\\
$E_\mathrm{T}^\mathrm{miss} > $ 240 GeV & $ 2.3\times 10^{-2} $ & $ 2.1^{+0.04}_{-0.04}\times 10^{-2} $ (stat) ${}^{+ 3.5\%}_{-3.1\%}$ (syst)\\
$m_{\mathrm{bb}}$ $\in$ [100,140] GeV & $ 1.7\times 10^{-2} $ & $ 1.8^{+0.04}_{-0.04}\times 10^{-2} $ (stat) ${}^{+ 3.5\%}_{-3.3\%}$ (syst)\\
$m_{\mathrm{T}}$ $>$ 100 GeV & $ 1.2\times 10^{-2} $ & $ 1.4^{+0.04}_{-0.04}\times 10^{-2} $ (stat) ${}^{+ 3.4\%}_{-3.4\%}$ (syst)\\ \hline
  $m_{\mathrm{CT}}$ $>$ 180 GeV & $ 9.7\times 10^{-3} $ & $ 9.7^{+0.31}_{-0.31}\times 10^{-3} $ (stat) ${}^{+ 5.0\%}_{-4.4\%}$ (syst)\\ \hline
  $m_{\mathrm{CT}}$ $\in$ [180,230] GeV & $ 1.0\times 10^{-3} $ & $ 8.4^{+0.90}_{-0.90}\times 10^{-4} $ (stat) ${}^{+ 4.5\%}_{-4.6\%}$ (syst)\\ \hline
  $m_{\mathrm{CT}}$ $\in$ [230,280] GeV & $ 1.7\times 10^{-3} $ & $ 9.5^{+0.96}_{-0.96}\times 10^{-4} $ (stat) ${}^{+ 6.5\%}_{-6.9\%}$ (syst)\\ \hline
  $m_{\mathrm{CT}}$ $>$ 280 GeV & $ 1.8\times 10^{-3} $ & $ 2.3^{+0.15}_{-0.15}\times 10^{-3} $ (stat) ${}^{+ 9.2\%}_{-5.7\%}$ (syst)\\
\hline\hline\end{tabular}
\caption{\label{TAB:LM}ATLAS-SUSY-2019-08, Signal regions LM for parameter point $m_{\tilde{\chi}_2^0}/m_{\tilde{\chi}_1^\pm}, m_{\tilde{\chi}_1^0} = (300 ,75 ) $ GeV. The final lines correspond to regions LMdisc,LMlow,LMmed and LMhigh respectively.}
\end{table}

\begin{table}\small\centering
\begin{tabular}{|c||c|c||} \hline\hline
Cut & ATLAS & HackAnalysis \\ \hline\hline
All weighted events & $ 1.0 $ & $ 1.0^{+0.00}_{-0.00} $ (stat) ${}^{+ 12.4\%}_{-9.4\%}$ (syst)\\
$N_{\mathrm{jets,25}} \geq 2$ & $ 8.6\times 10^{-1} $ & $ 8.9^{+0.01}_{-0.01}\times 10^{-1} $ (stat) ${}^{+ 0.7\%}_{-0.7\%}$ (syst)\\
1 signal lepton & $ 7.6\times 10^{-1} $ & $ 7.7^{+0.02}_{-0.02}\times 10^{-1} $ (stat) ${}^{+ 0.7\%}_{-0.9\%}$ (syst)\\
Second baseline lepton veto & $ 7.4\times 10^{-1} $ & $ 7.7^{+0.02}_{-0.02}\times 10^{-1} $ (stat) ${}^{+ 0.7\%}_{-0.9\%}$ (syst)\\
$m_{\mathrm{T}} > $ 50 GeV & $ 6.5\times 10^{-1} $ & $ 6.8^{+0.02}_{-0.02}\times 10^{-1} $ (stat) ${}^{+ 0.7\%}_{-0.9\%}$ (syst)\\
$E_\mathrm{T}^\mathrm{miss} > $ 180 GeV & $ 4.6\times 10^{-1} $ & $ 4.8^{+0.02}_{-0.02}\times 10^{-1} $ (stat) ${}^{+ 1.1\%}_{-1.3\%}$ (syst)\\
$N_{\mathrm{jets}} \leq 3$ & $ 3.8\times 10^{-1} $ & $ 3.6^{+0.02}_{-0.02}\times 10^{-1} $ (stat) ${}^{+ 1.8\%}_{-2.1\%}$ (syst)\\
$N_{\mathrm{b-jets}}=2$ & $ 1.6\times 10^{-1} $ & $ 1.7^{+0.01}_{-0.01}\times 10^{-1} $ (stat) ${}^{+ 3.0\%}_{-3.3\%}$ (syst)\\
$m_{\mathrm{bb}}$ $>$ 50 GeV & $ 1.6\times 10^{-1} $ & $ 1.7^{+0.01}_{-0.01}\times 10^{-1} $ (stat) ${}^{+ 3.0\%}_{-3.3\%}$ (syst)\\
$E_\mathrm{T}^\mathrm{miss} > $ 240 GeV & $ 1.1\times 10^{-1} $ & $ 1.2^{+0.01}_{-0.01}\times 10^{-1} $ (stat) ${}^{+ 2.7\%}_{-3.1\%}$ (syst)\\
$m_{\mathrm{bb}}$ $\in$ [100,140] GeV & $ 8.6\times 10^{-2} $ & $ 1.0^{+0.01}_{-0.01}\times 10^{-1} $ (stat) ${}^{+ 3.2\%}_{-3.5\%}$ (syst)\\
$m_{\mathrm{T}}$ $>$ 160 GeV & $ 6.6\times 10^{-2} $ & $ 8.3^{+0.11}_{-0.11}\times 10^{-2} $ (stat) ${}^{+ 3.2\%}_{-3.6\%}$ (syst)\\ \hline
  $m_{\mathrm{CT}}$ $>$ 180 GeV & $ 4.8\times 10^{-2} $ & $ 6.0^{+0.09}_{-0.09}\times 10^{-2} $ (stat) ${}^{+ 3.5\%}_{-3.8\%}$ (syst)\\ \hline
  $m_{\mathrm{CT}}$ $\in$ [180,230] GeV & $ 3.3\times 10^{-3} $ & $ 2.7^{+0.20}_{-0.20}\times 10^{-3} $ (stat) ${}^{+ 4.0\%}_{-4.2\%}$ (syst)\\ \hline
  $m_{\mathrm{CT}}$ $\in$ [230,280] GeV & $ 4.3\times 10^{-3} $ & $ 3.0^{+0.21}_{-0.21}\times 10^{-3} $ (stat) ${}^{+ 5.2\%}_{-5.0\%}$ (syst)\\ \hline
  $m_{\mathrm{CT}}$ $>$ 280 GeV & $ 6.8\times 10^{-3} $ & $ 8.5^{+0.36}_{-0.36}\times 10^{-3} $ (stat) ${}^{+ 4.5\%}_{-3.9\%}$ (syst)\\
\hline\hline\end{tabular}
\caption{\label{TAB:MM}ATLAS-SUSY-2019-08, Signal regions MM for parameter point $m_{\tilde{\chi}_2^0}/m_{\tilde{\chi}_1^\pm}, m_{\tilde{\chi}_1^0} = (500 ,0 ) $ GeV. The final lines correspond to regions MMdisc,MMlow,MMmed and MMhigh respectively.}
\end{table}

\begin{table}\small\centering
\begin{tabular}{|c||c|c||} \hline\hline
Cut & ATLAS & HackAnalysis \\ \hline\hline
All weighted events & $ 1.0 $ & $ 1.0^{+0.00}_{-0.00} $ (stat) ${}^{+ 16.1\%}_{-10.4\%}$ (syst)\\
$N_{\mathrm{jets,25}} \geq 2$ & $ 8.8\times 10^{-1} $ & $ 9.0^{+0.01}_{-0.01}\times 10^{-1} $ (stat) ${}^{+ 0.4\%}_{-0.5\%}$ (syst)\\
1 signal lepton & $ 7.9\times 10^{-1} $ & $ 7.9^{+0.02}_{-0.02}\times 10^{-1} $ (stat) ${}^{+ 0.4\%}_{-0.6\%}$ (syst)\\
Second baseline lepton veto & $ 7.6\times 10^{-1} $ & $ 7.9^{+0.02}_{-0.02}\times 10^{-1} $ (stat) ${}^{+ 0.4\%}_{-0.6\%}$ (syst)\\
$m_{\mathrm{T}} > $ 50 GeV & $ 7.0\times 10^{-1} $ & $ 7.3^{+0.02}_{-0.02}\times 10^{-1} $ (stat) ${}^{+ 0.4\%}_{-0.6\%}$ (syst)\\
$E_\mathrm{T}^\mathrm{miss} > $ 180 GeV & $ 6.0\times 10^{-1} $ & $ 6.2^{+0.02}_{-0.02}\times 10^{-1} $ (stat) ${}^{+ 0.5\%}_{-0.7\%}$ (syst)\\
$N_{\mathrm{jets}} \leq 3$ & $ 5.0\times 10^{-1} $ & $ 4.6^{+0.02}_{-0.02}\times 10^{-1} $ (stat) ${}^{+ 1.8\%}_{-1.8\%}$ (syst)\\
$N_{\mathrm{b-jets}}=2$ & $ 2.2\times 10^{-1} $ & $ 2.2^{+0.02}_{-0.02}\times 10^{-1} $ (stat) ${}^{+ 2.9\%}_{-2.6\%}$ (syst)\\
$m_{\mathrm{bb}}$ $>$ 50 GeV & $ 2.2\times 10^{-1} $ & $ 2.2^{+0.02}_{-0.02}\times 10^{-1} $ (stat) ${}^{+ 2.9\%}_{-2.6\%}$ (syst)\\
$E_\mathrm{T}^\mathrm{miss} > $ 240 GeV & $ 1.9\times 10^{-1} $ & $ 1.9^{+0.02}_{-0.02}\times 10^{-1} $ (stat) ${}^{+ 2.9\%}_{-2.6\%}$ (syst)\\
$m_{\mathrm{bb}}$ $\in$ [100,140] GeV & $ 1.4\times 10^{-1} $ & $ 1.6^{+0.01}_{-0.01}\times 10^{-1} $ (stat) ${}^{+ 3.0\%}_{-2.7\%}$ (syst)\\
$m_{\mathrm{\ell,b_1}}$ $>$ 120 GeV & $ 1.3\times 10^{-1} $ & $ 1.5^{+0.01}_{-0.01}\times 10^{-1} $ (stat) ${}^{+ 3.0\%}_{-2.7\%}$ (syst)\\
$m_{\mathrm{T}}$ $>$ 240 GeV & $ 9.6\times 10^{-2} $ & $ 1.1^{+0.01}_{-0.01}\times 10^{-1} $ (stat) ${}^{+ 3.2\%}_{-2.8\%}$ (syst)\\ \hline
  $m_{\mathrm{CT}}$ $>$ 180 GeV & $ 8.3\times 10^{-2} $ & $ 9.4^{+0.12}_{-0.12}\times 10^{-2} $ (stat) ${}^{+ 3.4\%}_{-2.8\%}$ (syst)\\ \hline
  $m_{\mathrm{CT}}$ $\in$ [180,230] GeV & $ 1.6\times 10^{-2} $ & $ 1.5^{+0.05}_{-0.05}\times 10^{-2} $ (stat) ${}^{+ 4.4\%}_{-3.3\%}$ (syst)\\ \hline
  $m_{\mathrm{CT}}$ $\in$ [230,280] GeV & $ 1.8\times 10^{-2} $ & $ 1.7^{+0.05}_{-0.05}\times 10^{-2} $ (stat) ${}^{+ 4.5\%}_{-2.9\%}$ (syst)\\ \hline
  $m_{\mathrm{CT}}$ $>$ 280 GeV & $ 5.0\times 10^{-2} $ & $ 6.2^{+0.10}_{-0.10}\times 10^{-2} $ (stat) ${}^{+ 2.9\%}_{-2.8\%}$ (syst)\\  
\hline\hline\end{tabular}
\caption{\label{TAB:HM}ATLAS-SUSY-2019-08, Signal regions HM for parameter point $m_{\tilde{\chi}_2^0}/m_{\tilde{\chi}_1^\pm}, m_{\tilde{\chi}_1^0} = (750 ,100 ) $ GeV. The final lines correspond to regions HMdisc,HMlow,HMmed and HMhigh respectively.} 
\end{table}

\begin{table}[!h]\centering\small
\begin{tabular}{|c||c|c||} \hline\hline
Cut & ATLAS & HackAnalysis \\ \hline\hline
Initial number of events ($\mathcal{L}\times\sigma$) & $ 1.0 $ & $ 1.0 $ \\
Initial number of events ($\mathcal{L}\times\sigma_{\geq 1 \textrm{jet}}$) & $ 2.3\times 10^{-1} $ & $ 5.0\times 10^{-1} $ \\
$E_{\text{T}}^{\text{miss}}$ trigger & $ 4.8\times 10^{-2} $ & $ 1.2\times 10^{-1} $ \\
1 lepton and $\geq$ 1 track & $ 1.7\times 10^{-2} $ & $ 2.8\times 10^{-3} $ \\
veto $3 \text{GeV} < m_{\ell\ell} < 3.2 \text{GeV}$ & $ 1.7\times 10^{-2} $ & $ 2.6\times 10^{-3} $ \\
${E_{\mathrm{T}}^{\mathrm{miss}}} > 200$ GeV & $ 7.1\times 10^{-3} $ & $ 8.8\times 10^{-4} $ \\
${\min(\Delta\phi(\mathrm{any~jet},\mathbf{p}_{\mathrm{T}}^{\mathrm{miss}}))} > 0.4$ & $ 6.7\times 10^{-3} $ & $ 8.0\times 10^{-4} $ \\
${\Delta\phi(j_1,\mathbf{p}_{\mathrm{T}}^{\mathrm{miss}})} > 2.0$ & $ 6.7\times 10^{-3} $ & $ 7.8\times 10^{-4} $ \\
$0.5 < m_{\ell \text{track}} < 5$ GeV & $ 6.0\times 10^{-4} $ & $ 7.7\times 10^{-4} $ \\
$\Delta R_{\ell \text{track}} >$ 0.05 & $ 6.0\times 10^{-4} $ & $ 7.7\times 10^{-4} $ \\
Number of jets $\geq$ 1 & $ 6.0\times 10^{-4} $ & $ 7.7\times 10^{-4} $ \\
Leading jet $p_{\mathrm{T}}>100$ GeV & $ 6.0\times 10^{-4} $ & $ 7.5\times 10^{-4} $ \\
${E_{\mathrm{T}}^{\mathrm{miss}}}/{H_{\mathrm{T}}^{\smash{\mathrm{lep}}}}>30$ & $ 3.4\times 10^{-4} $ & $ 4.5\times 10^{-4} $ \\
$\Delta R_{\ell \text{track}} <$ 1.5 & $ 3.3\times 10^{-4} $ & $ 4.4\times 10^{-4} $ \\
Lepton $p_{\mathrm{T}}<10$ GeV & $ 3.1\times 10^{-4} $ & $ 4.3\times 10^{-4} $ \\
Track $p_{\mathrm{T}}<5$ GeV & $ 3.1\times 10^{-4} $ & $ 4.3\times 10^{-4} $ \\
$\Delta \phi(\ell, E_{\text{T}}^{\text{miss}}) <$ 1.0 & $ 2.3\times 10^{-4} $ & $ 3.1\times 10^{-4} $ \\
Opposite charge lepton-track pair & $ 2.1\times 10^{-4} $ & $ 3.1\times 10^{-4} $ \\
$m_{\ell \text{track}} < 5$ GeV & $ 2.1\times 10^{-4} $ & $ 3.1\times 10^{-4} $ \\
$m_{\ell \text{track}} < 4$ GeV & $ 2.0\times 10^{-4} $ & $ 2.6\times 10^{-4} $ \\
$m_{\ell \text{track}} < 3$ GeV & $ 1.4\times 10^{-4} $ & $ 2.0\times 10^{-4} $ \\
$m_{\ell \text{track}} < 2$ GeV & $ 6.1\times 10^{-5} $ & $ 9.4\times 10^{-5} $ \\
$m_{\ell \text{track}} < 1.5$ GeV & $ 2.2\times 10^{-5} $ & $ 4.3\times 10^{-5} $ \\
$m_{\ell \text{track}} < 1$ GeV & $ 0.0 $ & $ 0.0 $ \\
\hline\hline\end{tabular}
\caption{\label{TAB:1L1T} Signal region comparison for ATLAS-SUSY-2018-16 signal region with one lepton and one track ({\tt SR\_E\_lT}).}
\end{table}

\begin{table}[!h]\centering\small
\begin{tabular}{|c||c|c||} \hline\hline
Cut & ATLAS & HackAnalysis \\ \hline\hline
Initial number of events ($\mathcal{L}\times\sigma$) & $ 1.0 $ & $ 1.0 $ \\
Initial number of events ($\mathcal{L}\times\sigma_{\geq 1 \textrm{jet}}$) & $ 2.3\times 10^{-1} $ & $ 5.0\times 10^{-1} $ \\
$E_{\text{T}}^{\text{miss}}$ trigger & $ 2.8\times 10^{-2} $ & $ 1.2\times 10^{-1} $ \\
2 leptons & $ 4.2\times 10^{-3} $ & $ 6.1\times 10^{-3} $ \\
veto $3 \text{GeV} < m_{\ell\ell} < 3.2 \text{GeV}$ & $ 3.9\times 10^{-3} $ & $ 5.7\times 10^{-3} $ \\
${\min(\Delta\phi(\mathrm{any~jet},\mathbf{p}_{\mathrm{T}}^{\mathrm{miss}}))} > 0.4$ & $ 3.8\times 10^{-3} $ & $ 5.3\times 10^{-3} $ \\
${\Delta\phi(j_1,\mathbf{p}_{\mathrm{T}}^{\mathrm{miss}})} > 2.0$ & $ 3.7\times 10^{-3} $ & $ 5.2\times 10^{-3} $ \\
$1 < {m_{\ell\ell}} < 60$ GeV & $ 3.3\times 10^{-3} $ & $ 4.0\times 10^{-3} $ \\
$\Delta R_{ee} >$ 0.30, $\Delta R_{\mu\mu} >$ 0.05, $\Delta R_{e\mu} >$ 0.20 & $ 2.9\times 10^{-3} $ & $ 4.0\times 10^{-3} $ \\
Leading lepton $p_{\mathrm{T}}$ $>$ 5 GeV & $ 2.4\times 10^{-3} $ & $ 3.3\times 10^{-3} $ \\
Number of jets $\geq$ 1 & $ 2.3\times 10^{-3} $ & $ 3.3\times 10^{-3} $ \\
Leading jet $p_{\mathrm{T}}$ $>$100 GeV & $ 2.1\times 10^{-3} $ & $ 2.5\times 10^{-3} $ \\
Number of b-tagged jets = 0 & $ 1.8\times 10^{-3} $ & $ 2.2\times 10^{-3} $ \\
$m_{\tau\tau}$ $<$ 0 or $>$ 160 GeV & $ 1.5\times 10^{-3} $ & $ 1.9\times 10^{-3} $ \\
ee or $\mu\mu$ & $ 1.5\times 10^{-3} $ & $ 1.9\times 10^{-3} $ \\
${m_{\mathrm{T}}^{\ell_1}} < 60$ GeV & $ 1.3\times 10^{-3} $ & $ 1.6\times 10^{-3} $ \\
${E_{\mathrm{T}}^{\mathrm{miss}}}$ $>$ 200 & $ 6.5\times 10^{-4} $ & $ 8.9\times 10^{-4} $ \\
$\max(0.85, 0.98-0.02\times m_{\ell\ell}) < {R_{\mathrm{ISR}}} < 1.0$ & $ 4.9\times 10^{-4} $ & $ 5.6\times 10^{-4} $ \\
sub-leading lepton $p_{\mathrm{T}} > \min(10, 2 + {m_{\ell\ell}}/3)$ & $ 4.7\times 10^{-4} $ & $ 5.5\times 10^{-4} $ \\
$m_{\ell\ell} < 60$ GeV & $ 4.7\times 10^{-4} $ & $ 5.5\times 10^{-4} $ \\
$m_{\ell\ell} < 40$ GeV & $ 4.7\times 10^{-4} $ & $ 5.5\times 10^{-4} $ \\
$m_{\ell\ell} < 30$ GeV & $ 4.7\times 10^{-4} $ & $ 5.5\times 10^{-4} $ \\
$m_{\ell\ell} < 20$ GeV & $ 4.7\times 10^{-4} $ & $ 5.5\times 10^{-4} $ \\
$m_{\ell\ell} < 10$ GeV & $ 4.7\times 10^{-4} $ & $ 5.5\times 10^{-4} $ \\
$m_{\ell\ell} < 5$ GeV & $ 4.7\times 10^{-4} $ & $ 5.5\times 10^{-4} $ \\
$m_{\ell\ell} < 3$ GeV & $ 3.2\times 10^{-4} $ & $ 3.6\times 10^{-4} $ \\
$m_{\ell\ell} < 2$ GeV & $ 1.3\times 10^{-4} $ & $ 1.5\times 10^{-4} $ \\
\hline\hline\end{tabular}
\caption{\label{TAB:SREhigh} Signal region comparison for ATLAS-SUSY-2018-16 signal region {\tt SR\_E\_high}.}
\end{table}

\begin{table}[!h]\centering\small
\begin{tabular}{|c||c|c||} \hline\hline
Cut & ATLAS & HackAnalysis \\ \hline\hline
Initial number of events ($\mathcal{L}\times\sigma$) & $ 1.0 $ & $ 1.0 $ \\
Initial number of events ($\mathcal{L}\times\sigma_{\geq 1 \textrm{jet}}$) & $ 2.3\times 10^{-1} $ & $ 5.0\times 10^{-1} $ \\
$E_{\text{T}}^{\text{miss}}$ trigger & $ 2.8\times 10^{-2} $ & $ 1.2\times 10^{-1} $ \\
2 leptons & $ 4.2\times 10^{-3} $ & $ 6.1\times 10^{-3} $ \\
veto $3 \text{GeV} < m_{\ell\ell} < 3.2 \text{GeV}$ & $ 3.9\times 10^{-3} $ & $ 5.7\times 10^{-3} $ \\
${\min(\Delta\phi(\mathrm{any~jet},\mathbf{p}_{\mathrm{T}}^{\mathrm{miss}}))} > 0.4$ & $ 3.8\times 10^{-3} $ & $ 5.3\times 10^{-3} $ \\
${\Delta\phi(j_1,\mathbf{p}_{\mathrm{T}}^{\mathrm{miss}})} > 2.0$ & $ 3.7\times 10^{-3} $ & $ 5.2\times 10^{-3} $ \\
$1 < {m_{\ell\ell}} < 60$ GeV & $ 3.3\times 10^{-3} $ & $ 4.0\times 10^{-3} $ \\
$\Delta R_{ee} >$ 0.30, $\Delta R_{\mu\mu} >$ 0.05, $\Delta R_{e\mu} >$ 0.20 & $ 2.9\times 10^{-3} $ & $ 4.0\times 10^{-3} $ \\
Leading lepton $p_{\mathrm{T}}$ $>$ 5 GeV & $ 2.4\times 10^{-3} $ & $ 3.3\times 10^{-3} $ \\
Number of jets $\geq$ 1 & $ 2.3\times 10^{-3} $ & $ 3.3\times 10^{-3} $ \\
Leading jet $p_{\mathrm{T}}$ $>$100 GeV & $ 2.1\times 10^{-3} $ & $ 2.5\times 10^{-3} $ \\
Number of b-tagged jets = 0 & $ 1.8\times 10^{-3} $ & $ 2.2\times 10^{-3} $ \\
$m_{\tau\tau}$ $<$ 0 or $>$ 160 GeV & $ 1.5\times 10^{-3} $ & $ 1.9\times 10^{-3} $ \\
ee or $\mu\mu$ & $ 1.5\times 10^{-3} $ & $ 1.9\times 10^{-3} $ \\
$120 < {E_{\mathrm{T}}^{\mathrm{miss}}} < 200$ GeV & $ 5.9\times 10^{-4} $ & $ 6.9\times 10^{-4} $ \\
${E_{\mathrm{T}}^{\mathrm{miss}}}/{H_{\mathrm{T}}^{\smash{\mathrm{lep}}}} < 10.0$ & $ 1.7\times 10^{-4} $ & $ 2.0\times 10^{-4} $ \\
$ 0.8 < {R_{\mathrm{ISR}}} < 1.0$ & $ 1.3\times 10^{-4} $ & $ 1.4\times 10^{-4} $ \\
sub-leading lepton $p_{\mathrm{T}} > \min(5 + {m_{\ell\ell}}/4)$ & $ 8.5\times 10^{-5} $ & $ 9.6\times 10^{-5} $ \\
$10 < {m_{\mathrm{T}}^{\ell_1}} < 60$ GeV & $ 6.1\times 10^{-5} $ & $ 7.1\times 10^{-5} $ \\
$m_{\ell\ell} < 60$ GeV & $ 6.1\times 10^{-5} $ & $ 7.1\times 10^{-5} $ \\
$m_{\ell\ell} < 40$ GeV & $ 6.1\times 10^{-5} $ & $ 7.1\times 10^{-5} $ \\
$m_{\ell\ell} < 30$ GeV & $ 6.1\times 10^{-5} $ & $ 7.1\times 10^{-5} $ \\
$m_{\ell\ell} < 20$ GeV & $ 6.1\times 10^{-5} $ & $ 7.1\times 10^{-5} $ \\
$m_{\ell\ell} < 10$ GeV & $ 6.1\times 10^{-5} $ & $ 7.1\times 10^{-5} $ \\
$m_{\ell\ell} < 5$ GeV & $ 6.0\times 10^{-5} $ & $ 7.1\times 10^{-5} $ \\
$m_{\ell\ell} < 3$ GeV & $ 4.5\times 10^{-5} $ & $ 5.0\times 10^{-5} $ \\
$m_{\ell\ell} < 2$ GeV & $ 2.3\times 10^{-5} $ & $ 2.6\times 10^{-5} $ \\
\hline\hline\end{tabular}
\caption{\label{TAB:SRElow} Signal region comparison for ATLAS-SUSY-2018-16 signal region {\tt SR\_E\_low}.}
\end{table}

\begin{table}[!h]\centering\small
\begin{tabular}{|c||c|c||} \hline\hline
Cut & ATLAS & HackAnalysis \\ \hline\hline
Initial number of events ($\mathcal{L}\times\sigma$) & $ 1.0 $ & $ 1.0 $ \\
Initial number of events ($\mathcal{L}\times\sigma_{\geq 1 \textrm{jet}}$) & $ 2.3\times 10^{-1} $ & $ 5.0\times 10^{-1} $ \\
$E_{\text{T}}^{\text{miss}}$ trigger & $ 2.8\times 10^{-2} $ & $ 1.2\times 10^{-1} $ \\
2 leptons & $ 4.2\times 10^{-3} $ & $ 6.1\times 10^{-3} $ \\
veto $3 \text{GeV} < m_{\ell\ell} < 3.2 \text{GeV}$ & $ 3.9\times 10^{-3} $ & $ 5.7\times 10^{-3} $ \\
${\min(\Delta\phi(\mathrm{any~jet},\mathbf{p}_{\mathrm{T}}^{\mathrm{miss}}))} > 0.4$ & $ 3.8\times 10^{-3} $ & $ 5.3\times 10^{-3} $ \\
${\Delta\phi(j_1,\mathbf{p}_{\mathrm{T}}^{\mathrm{miss}})} > 2.0$ & $ 3.7\times 10^{-3} $ & $ 5.2\times 10^{-3} $ \\
$1 < {m_{\ell\ell}} < 60$ GeV & $ 3.3\times 10^{-3} $ & $ 4.0\times 10^{-3} $ \\
$\Delta R_{ee} >$ 0.30, $\Delta R_{\mu\mu} >$ 0.05, $\Delta R_{e\mu} >$ 0.20 & $ 2.9\times 10^{-3} $ & $ 4.0\times 10^{-3} $ \\
Leading lepton $p_{\mathrm{T}}$ $>$ 5 GeV & $ 2.4\times 10^{-3} $ & $ 3.3\times 10^{-3} $ \\
Number of jets $\geq$ 1 & $ 2.3\times 10^{-3} $ & $ 3.3\times 10^{-3} $ \\
Leading jet $p_{\mathrm{T}}$ $>$100 GeV & $ 2.1\times 10^{-3} $ & $ 2.5\times 10^{-3} $ \\
Number of b-tagged jets = 0 & $ 1.8\times 10^{-3} $ & $ 2.2\times 10^{-3} $ \\
$m_{\tau\tau}$ $<$ 0 or $>$ 160 GeV & $ 1.5\times 10^{-3} $ & $ 1.9\times 10^{-3} $ \\
ee or $\mu\mu$ & $ 1.5\times 10^{-3} $ & $ 1.9\times 10^{-3} $ \\
$120 < {E_{\mathrm{T}}^{\mathrm{miss}}} < 200$ GeV & $ 5.9\times 10^{-4} $ & $ 6.9\times 10^{-4} $ \\
${E_{\mathrm{T}}^{\mathrm{miss}}}/{H_{\mathrm{T}}^{\smash{\mathrm{lep}}}} > 10$ & $ 4.2\times 10^{-4} $ & $ 4.8\times 10^{-4} $ \\
${M_{\mathrm{T}}^{\mathrm{S}}} <50$ GeV & $ 2.6\times 10^{-4} $ & $ 2.5\times 10^{-4} $ \\
$m_{\ell\ell} < 30$ GeV & $ 2.6\times 10^{-4} $ & $ 2.5\times 10^{-4} $ \\
$m_{\ell\ell} < 20$ GeV & $ 2.6\times 10^{-4} $ & $ 2.5\times 10^{-4} $ \\
$m_{\ell\ell} < 10$ GeV & $ 2.6\times 10^{-4} $ & $ 2.5\times 10^{-4} $ \\
$m_{\ell\ell} < 5$ GeV & $ 2.6\times 10^{-4} $ & $ 2.5\times 10^{-4} $ \\
$m_{\ell\ell} < 3$ GeV & $ 2.0\times 10^{-4} $ & $ 1.8\times 10^{-4} $ \\
$m_{\ell\ell} < 2$ GeV & $ 9.8\times 10^{-5} $ & $ 9.3\times 10^{-5} $ \\
\hline\hline\end{tabular}
\caption{\label{TAB:SREmed} Signal region comparison for ATLAS-SUSY-2018-16 signal region {\tt SR\_E\_med}.}
\end{table}



\bibliographystyle{h-physrev}
\bibliography{lit}

\end{document}